\documentclass[sigconf]{acmart}

\renewcommand\footnotetextcopyrightpermission[1]{} % removes footnote with conference information in first column

\usepackage{epsfig}
\usepackage{graphicx}
\usepackage{amsmath}
\usepackage{amssymb}

\usepackage{enumitem}
\usepackage{tablefootnote}
\usepackage[linesnumbered,titlenumbered,ruled,vlined,commentsnumbered,noend]{algorithm2e}
\usepackage{stfloats}
\usepackage{makecell}
\usepackage{colortbl}
\definecolor{linen}{rgb}{0.96, 0.94, 0.93}
\definecolor{lightcyan}{rgb}{0.88, 1.0, 1.0}
\definecolor{lightyellow}{rgb}{1.0, 1.0, 0.88}
\definecolor{azure}{rgb}{0.94, 1.0, 1.0}

\usepackage[font=footnotesize,labelfont=bf]{caption}

\usepackage{multirow}
\usepackage{tabularx}
\usepackage{dsfont}
\usepackage{url}
\usepackage[tight]{subfigure}
\usepackage{color}
\usepackage{subfigure}
\usepackage{amsthm}
\usepackage[export]{adjustbox}
\usepackage{comment}

\usepackage[utf8]{inputenc} % allow utf-8 input
\usepackage[T1]{fontenc}    % use 8-bit T1 fonts
\usepackage{url}            % simple URL typesetting
\usepackage{booktabs}       % professional-quality tables
\usepackage{amsfonts}       % blackboard math symbols
\usepackage{nicefrac}       % compact symbols for 1/2, etc.
\usepackage{microtype}      % microtypography

\definecolor{dg}{rgb}{0,0.694,0.298}
\definecolor{purple}{rgb}{0.4,0.176,0.569}
\usepackage{pifont}% http://ctan.org/pkg/pifont
\usepackage{xspace}
\makeatletter
\DeclareRobustCommand\onedot{\futurelet\@let@token\@onedot}
\def\@onedot{\ifx\@let@token.\else.\null\fi\xspace}
\def\eg{\emph{e.g}\onedot} 
\def\ie{\emph{i.e}\onedot} 
 
\def\etc{\emph{etc}\onedot} 
 
\def\etal{\emph{et al}\onedot}
\makeatother

\AtBeginDocument{%
  \providecommand\BibTeX{{%
    \normalfont B\kern-0.5em{\scshape i\kern-0.25em b}\kern-0.8em\TeX}}}

\begin{document}

% \title{\emph{Free Fine-tuning}: A Practical yet Robust Watermarking Scheme for Deep Neural Networks}
\title{\emph{Free Fine-tuning}: A Plug-and-Play Watermarking Scheme for Deep Neural Networks}

\author{Run Wang}
\authornote{School of Cyber Science and Engineering, Wuhan University}
\affiliation{
  \institution{Wuhan University}
}

\author{Jixing Ren}
\authornotemark[1]
\affiliation{
  \institution{Wuhan University}
}

\author{Boheng Li}
\authornotemark[1]
\affiliation{
  \institution{Wuhan University}
}

\author{Tianyi She}
\authornotemark[1]
\affiliation{
  \institution{Wuhan University}
}

\author{Chenhao Lin}
% \authornotemark[2]
\affiliation{
  \institution{Xi'an Jiaotong University}
}

\author{Liming Fang}
\affiliation{
  \institution{Nanjing University of Aeronautics and Astronautics}
}

\author{Jing Chen}
\authornotemark[1]
\affiliation{
  \institution{Wuhan University}
}

\author{Chao Shen}
% \authornotemark[2]
\affiliation{
  \institution{Xi'an Jiaotong University}
}

\author{Lina Wang}
\authornotemark[1]
\affiliation{
  \institution{Wuhan University}
}

% \affiliation{%

% \affiliation{%

\renewcommand{\shortauthors}{Trovato and Tobin, et al.}

\begin{abstract}

Watermarking has been widely adopted for protecting the intellectual property (IP) of Deep Neural Networks (DNN) to defend the unauthorized distribution. Unfortunately, the popular data-poisoning DNN watermarking scheme relies on target model fine-tuning to embed watermarks, which limits its practical applications in tackling real-world tasks. Specifically, the learning of watermarks via tedious model fine-tuning on a poisoned dataset (carefully-crafted sample-label pairs) is not efficient in tackling the tasks on challenging datasets and production-level DNN model protection.

To address the aforementioned limitations, in this paper, we propose a plug-and-play watermarking scheme for DNN models by injecting an independent proprietary model into the target model to serve the watermark embedding and ownership verification. In contrast to the prior studies, our proposed method by incorporating a proprietary model is free of target model fine-tuning without involving any parameters update of the target model, thus the fidelity is well preserved. Furthermore, our method is scaleable to challenging datasets, large production-level models, and diverse tasks (\eg{}, speaker recognition). Experimental results on real-world challenging datasets (\eg{}, ImageNet) and real-world DNN models demonstrated its effectiveness, fidelity \textit{w.r.t.} the functionality preserving of the target model, robustness against popular watermark removal attacks (\ie{}, fine-tuning attack, pruning, input preprocessing), and the plug-and-play deployment. Our proposed watermarking scheme also outperforms the two competitive baselines in terms of fidelity preserving and robustness against watermark removal attacks. Our research findings reveal that model fine-tuning with poisoned data is not prepared for the IP protection of DNN models deployed in real-world tasks and poses a new research direction toward a more thorough understanding and investigation of adopting the proprietary model for DNN watermarking. The source code and models are available at \href{https://github.com/AntigoneRandy/PTYNet/}{\textcolor{blue}{https://github.com/AntigoneRandy/PTYNet.}}

\end{abstract}

% \ccsdesc[500]{Information systems~Multimedia information systems}


% \keywords{Disrupting watermark, shallow reconstruction, face recognition}



% %   \includegraphics[width=\textwidth]{sampleteaser}

\maketitle

%% symbols
%% \etal \eg \ie \etc
%reference

\section{Introduction}\label{sec:intro}
%  1 page
% 为什么要做DNN IP protection
In the past decade, DNN has achieved tremendous success in many cutting-edge fields \cite{pouyanfar2018survey}, such as autonomous driving \cite{luo2017traffic}, genomics \cite{zou2019primer}. However, training powerful DNN models, especially the so-called foundation models \cite{bommasani2021opportunities}, requires a large amount of valuable data and is computationally expensive. According to a report\footnote{https://venturebeat.com/ai/ai-machine-learning-openai-gpt-3-size-isnt-everything}, the OpenAI costs more than \$12 million for training GPT-3 \cite{brown2020language}. Thus, a well-trained DNN model has high value to the owner. Recently, some large companies like Google, Meta sell commercial high-value models to users for offering paid services, which is becoming a lucrative business. Unfortunately, the high-value well-trained DNN models have the potential threat to be stolen or extracted by adversaries through various unimaginable manners \cite{krishna2019thieves} and pose the threat of unauthorized distribution. Thus, effective countermeasures should be devised for the IP protection of DNN models.

% 目前的研究方法存在weakness,以及解决好这个问题,面临的challenge在哪里
Recently, DNN watermarking is widely employed for the IP protection of DNN models~\cite{lukas2021sok,boenisch2020survey,chen2021you,li2021survey} by embedding designed watermarks into the target DNN model. The original idea of DNN watermarking borrows from the digital multimedia protection~\cite{petitcolas1999information} to embed identification signals into the multimedia without introducing obvious visual quality degradations. In general, the \emph{parameter-embedding} and \emph{data-poisoning} are two mainstream watermarking schemes \cite{guo2021finetuning,yang2021robust,liu2021watermarking,chen2022copy}. Noticeably, the parameter embedding watermarking scheme requires white-box access to the suspicious model which is not practical in the real-world scenario \cite{wang2021riga,kuribayashi2021white}. The data-poisoning watermarking scheme crafts a set of sample-label pairs (also called \textit{verification samples}) to enforce the DNN model memorizing them via carefully model fine-tuning. Thus, the data-poisoning watermarking scheme is the most promising technique, which works in a black-box setting and extracts the embedded watermarks for ownership verification by querying the suspicious model only~\cite{barni2021dnn,lukas2021sok}. Specifically, the owner determines the ownership by checking the consistency of the desired output label of verification samples and its real output label. Unfortunately, the existing widely adopted data-poisoning watermarking scheme suffers the following two key challenges in the IP protection of DNN models in practice.

\begin{itemize}[leftmargin=*]
\item \textbf{Suffering fidelity degradation via target model fine-tuning}. The model fine-tuning inevitably updates the target model's parameters and introduces performance degradation to the model's original functionality, especially tackling the real-world large dataset (\eg{}, ImageNet) and challenging tasks that call for extremely skilled fine-tuning \cite{boenisch2020survey,wang2021non,jia2021entangled}.
% Thus, the model fine-tuning limits the true application of data-poisoning watermarking scheme in protecting the IP of DNN models \cite{chen2021refit}. 
% (\eg{}reinforcement learning involving interactions \cite{ayoub2020model})
\item \textbf{High time-consume and computation resource-costing}. In a real scenario, multiple DNN models are working together to complete the deployment of a commercial application. However, the existing watermarking scheme involves the target model fine-tuning for all the intentionally protected models, even some of them share a similar architecture.
% \item \textbf{Not enough for tackling watermark removal attacks}. A recent study reveals that none of the data-poisoning watermarking schemes is robust against removal attacks~\cite{lukas2021sok}. The model could easily forget the embedded watermarks via fine-tuning when suffering removal attacks like model pruning, input preprocessing.

% The watermark embedding requires the target model fine-tuning on poisoned dataset, thus such model fine-tuning should be conducted for each intentionally protected model even models sharing similar architecture.

% Furthermore, the model fine-tuning costs computation resources and is time-consuming when the version update frequently in practice.
% \item \textbf{Poor transferabilities across diverse tasks}. The watermarking schemes developed are mostly focus on particular tasks, like image classification. However, it will costs human efforts since each task will require customized watermarking technique, especially for a large vendor has a mass of production-level DNN models.

% \item \textbf{The embedded samples with pattern are not stealthiness}. The sample for verifying the ownership with certain patterns could be easily detected \cite{wang2019neural}. 
\end{itemize}

% 我们准备怎么做，方法的insight在哪里，优点是什么
To address the aforementioned inevitable key challenges, recently, there have been some initial attempts to investigate the unique fingerprints as a kind of special watermarks for the IP protection of DNN models \cite{li2021novel,peng2022fingerprinting}, especially exploring the samples near the decision boundary, such as perturbing normal samples~\cite{le2020adversarial} and exploring out-of-distribution (OOD) samples~\cite{adi2018turning}. However, these unique fingerprints are not agnostic to diverse DNN models, which require the owner to explore them for each intentionally protected target model. Most of the time, the unique fingerprints could be easily investigated by attackers, which could be evaded via fine-tuning or sample preprocessing. In this paper, for the first time, we propose a novel DNN watermarking scheme by injecting a proprietary model for watermark embedding in an efficient manner without sacrificing the fidelity of the target model. Specifically, our method is model-agnostic and works in black-box settings without obtaining any knowledge of the suspicious model in verification.
% researchers approach the IP protection of DNN model by investigating the unique fingerprints as watermarks

\begin{figure*}[t]
\centering
\includegraphics[width=0.8\linewidth]{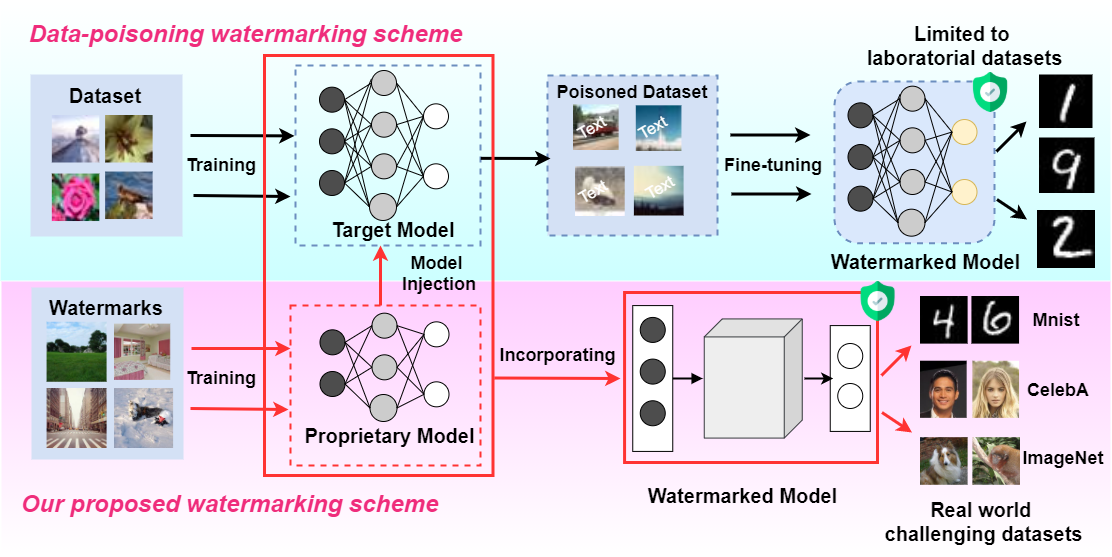}
% \vspace{-10pt}
\caption{An overview of the difference between prior data-poisoning watermarking scheme and our proposed watermarking scheme via injecting proprietary model for ownership verification. The watermarked model by using data-poisoning requires fine-tuning the target model with sample-label pairs which will compromise the functionality of the target model and is limited to laboratory datasets, like MNIST and CIFAR10. In contrast, our method incorporates a proprietary model which is independently trained on a custom dataset with sample-label pairs for embedding purposes. Due to free from fine-tuning, our method shows potentials for real-world challenging datasets (\eg{}, ImageNet) and recent popular vision transformer backbones \cite{lee2022mpvit,vit}.}
\label{fig:overview}
% \vspace{-10pt}
\end{figure*}

Our novel watermarking scheme via proprietary model is motivated by the simple idea from the principle of software designing in software engineering that \textit{modules require high cohesion and low coupling}. Thus, we devise a proprietary model for watermark embedding specifically without fine-tuning the target model to embed watermarks like the prior studies \cite{ruan2023intellectual,wang2022protecting}. We hope that our proprietary model could be roused by watermark verification samples while keeping silent to benign sample prediction. Figure~\ref{fig:overview} illustrates the comparison with the existing data-poisoning watermarking scheme and our proposed method. To comprehensively evaluate our proposed watermarking scheme, the experiments are conducted on real-world challenging datasets ImageNet with six popular DNN models and large-scale speaker identification dataset VoxCeleb1 with VGGVox for speaker recognition \cite{Nagrani17}. Additionally, we also evaluate the effectiveness on real-world production-level DNN models or state-of-the-art (SOTA) DNN backbones, such as ViT and commercial models offered by three vendors (\ie{}, Amazon, Google, Chooch) to provide online service. Experimental results show that our watermarking scheme preserves the model's functionality in nearly \textbf{100\%} confidence which significantly outperforms baselines, gives \textbf{100\%} accuracy in ownership verification, survives popular watermark removal attacks (\eg{}, model fine-tuning, network pruning, and input preprocessing) with competitive performance.
% \wang{our experiments are conducted on real-world challenging dataset ImageNet with six popular DNN models and large-scale speaker identification dataset VoxCeleb1 with VGGVox for speaker recognition \cite{Nagrani17}.} 

Our main contributions are summarized as follows:

\begin{itemize}[leftmargin=*]
\item We introduce a novel watermarking scheme by incorporating a proprietary model for watermark embedding and ownership verification. In contrast to the prior data-poisoning watermarking scheme, our proposed method is free of target model fine-tuning, which shows potential in tackling real tasks with production-level models on real-world challenging datasets.
% \item We propose an effective model patching method by injecting a tiny trojan module with multi-layer perceptron (MLP) to connect the first and last layer of the target model for verifying watermarks. Our injected tiny module is model-agnostic and has no impact to the original model's functionality.
% \item We propose the search-based and generation-based method for selecting trigger for image classification and conduct a comprehensive evaluation of proposed watermarking scheme in terms of \textit{effectiveness}, \textit{fidelity}, and \textit{robustness}, for the first time, on the real world ImageNet dataset with six different DNN models. Experimental results show its practicability in real scenarios.
\item We propose a generation-based method for crafting verification samples in a safety manner and conduct a comprehensive evaluation in terms of \textit{effectiveness}, \textit{fidelity}, \textit{robustness}, and \textit{efficiency}, for the first time, on the real-world ImageNet dataset and production-level DNN models. Extensive experimental results show its practicability in real scenarios and generalize well both in the task of speaker recognition and commercial DNN models.
% , commercial production-level DNN model with six different DNN models. Experimental results show its practicability in real scenarios and generalize well in the task of speaker recognition.
\item Our research findings imply a new research direction towards developing an independent proprietary model for watermark embedding by injecting it into the target model for IP protection, as opposed to fine-tuning the target model in prior studies. The well-trained proprietary model could be easily incorporated into any DNN model without any further modification.
\end{itemize}
% \vspace{-5pt}

%-------------------------------------------------------------------------
%-------------------------------------------------------------------------
\section{Related Work}\label{sec:related}
% \vspace{-5pt}
% 0.75 page
% Recently, we have witnessed the progress of DNN watermarking techniques and attacks to remove the embedded watermarks. This section briefly reviews the related work to our model patching schemes.

% In recent years, we have witnessed many research efforts and encouraging progress on developing powerful DNN watermarking techniques and exploring the vulnerabilities of these techniques to remove them. This section briefly reviews the related work to our proposed method.
\subsection{DNN Watermarking}
We systematize the existing DNN watermarking schemes into parameter-embedding and data-poisoning watermarking schemes based on whether the owner needs to access the suspicious model in ownership verification. 
% We systematize the existing DNN watermarking schemes into parameter-embedding \cite{rouhani2019deepsigns} and data-poisoning \cite{le2020adversarial} watermarking scheme based on whether the owner needs to access the suspicious model in ownership verification. 
% Next, we briefly review these two popular DNN watermarking schemes and present their shortcomings as well.

% We systematize the existing DNN watermarking schemes into parameter-embedding \cite{uchida2017embedding,rouhani2019deepsigns,chen2018deepmarks} and data-poisoning \cite{adi2018turning,zhang2018protecting,li2019prove,le2020adversarial} watermarking schemes based on whether the owner needs to access the suspicious model in ownership verification. Next, we briefly review these two popular DNN watermarking schemes and present their shortcomings as well.

\textbf{Parameter-embedding} watermarking scheme embeds watermarks into the target model's parameters \cite{uchida2017embedding,kuribayashi2021white} or the activations of hidden layers \cite{rouhani2019deepsigns,mehta2022aime}. \citet{uchida2017embedding} proposed embedding watermarks into the model parameters by using a parameter regularizer with a designed \emph{embedding} loss. DeepSigns~\cite{rouhani2019deepsigns} proposed an end-to-end watermarking embedding framework to embed watermarks into the activation maps in various layers. However, all of these parameter-embedding watermarking schemes require to access the model weights during verification (\ie, {white-box setting}), thus are not practical for real-world scenarios.
% DeepMarks~\cite{chen2018deepmarks} adopted a similar watermark embedding approach as in Uchinda.

\textbf{Data-poisoning} watermarking scheme crafts sample-label pairs as watermarks via model fine-tuning and verifies the watermarks by querying the model in the black-box setting \cite{li2022leveraging,lv2022svd,goldblum2022dataset}. The sample could be generated by blending certain patterns (called \emph{pattern-based}), perturbed on normal samples (called \emph{perturbation-based}), or drawn from other data sources, also known as OOD (called \emph{OOD-based}). For the pattern-based, \citet{zhang2018protecting} proposed a crafted watermark generation method by taking a subset of training images and adding meaningful content like a special string ``TEST" onto them. For the perturbation-based, \citet{le2020adversarial} leveraged adversarial examples as watermarks to obtain the samples nearby decision frontiers. For the OOD-based, \citet{zhang2018protecting} used handwritten image ``1" as the watermark in CIFAR10 dataset and assigned it a ``airplane" label. For ownership verification, if the protected model recognizes the handwritten image ``1" as ``airplane", the owner can claim the possession of this model. Some studies working on enhancing the model by modifying the architecture \cite{fan2019rethinking} or a few weights \cite{lao2022few} of the network for ownership verification or applicability authorization. \citet{fan2019rethinking} insert \emph{passport layers} into the model, where the model will perform badly when passport weights are not present. \citet{lao2022few} adjust a few weights of the model to embed watermarks for ownership verification.

% \Li{Model-enhancement} watermarking scheme modifies the architecture \cite{fan2019rethinking} or a few weights \cite{lao2022few} of the network for ownership verification or applicability authorization. Fan \etal{} \cite{fan2019rethinking} insert \emph{passport layers} into the model, where the model will perform badly when passport weights are not present. Lao \etal{} \cite{lao2022few} adjust few weights of the model to embed watermarks.

Unfortunately, the paradigm of the prior data-poisoning watermarking scheme needs target model fine-tuning for further ownership verification which suffers performance degradation in benign sample prediction and is time-consuming. To avoid these shortcomings, we develop a practical watermarking scheme by injecting a proprietary model into the target model in an efficient manner for ownership verification and fidelity preservation purpose.

\subsection{Watermark Removal Attack}
Studies are also working on exploring the vulnerabilities of DNN watermarking techniques to remove the watermarks by conducting model modification or input preprocessing.
% . The removal attack can be classified into

\textbf{Model modification} updates the model's parameters or modifies the model's architecture to remove the embedded watermarks~\cite{wang2019attacks, liu2018fine, chen2021refit,ong2021protecting}, such as network pruning or model fine-tuning. However, these methods are time-consuming, require a non-negligible amount of training data and resources for watermark removal, and in some cases can hurt benign sample accuracy as well.

\textbf{Input preprocessing} aims at corrupting the embedded watermark triggers at inference time by conducting various transformation techniques, such as input reconstruction, input smoothing, image scaling~\cite{guo2021finetuning}, relighting~\cite{removal} \etc{}. \citet{guo2021finetuning} proposed PST with a series of image transformation techniques (\eg{} scaling, embedding random imperceptible patterns, and spatial-level transformations) to remove watermarks blindly. A very recent work \cite{removal} introduced naturalness-aware relighting perturbations to mask the embedded watermark triggers, which achieved the SOTA performance in disrupting verification samples. Since input preprocessing techniques are usually watermark-scheme-agnostic, model-independent, and training data careless, it poses the biggest threats to the survival of DNN watermarks.

\section{Motivation and Design Insight}\label{sec:motivation}
\subsection{Background on Watermarking}
Watermarking is a promising technique to protect the IP of DNN. For a typical data-poisoning watermarking scheme, it requires sample-label pairs $\mathcal{K}$$=$$\{(x^n,y^n)\}^{N}_{n=1}$ to enforce the model could remember them via fine-tuning. Then, sample-label pairs could be leveraged to query the suspicious model for ownership verification. Here, we present the threat model of DNN watermarking.

\textbf{Threat Model}. Given a target model $\mathcal{M}$, the owner needs to protect the model $\mathcal{M}$ to avoid illegal distribution, while the adversary may obtain a suspicious model $\mathcal{M}^{s}$ under unauthorized distribution. Specifically, the owner could merely access the suspicious model $\mathcal{M}^{s}$ by sending inputs (also known as verification samples) to verify the output for further checking whether $\mathcal{M}^{s}$ is a copy of $\mathcal{M}$. However, the adversary may set a series of obstacles to prevent such ownership verification process. First, the model $\mathcal{M}$ could be modified intentionally, like fine-tuning and model pruning. Second, the input could be preprocessed to destroy a certain pattern in verification samples. Thus, a practical watermarking scheme should be robust against these watermark removal attacks and functionality preserving in benign sample prediction.
% \vspace{-5pt}

\subsection{Challenges to Practical Watermarking}
The existing studies mainly focus on how to improve the robustness of watermarking techniques to evade common watermark removal attacks. However, these studies are merely evaluating their performance in the laboratory scenario with simple datasets (\eg{}, CIFAR10, CIFAR100) on small-scale DNN models, where another important problem on \textit{whether these techniques can generalize to large real-world datasets} are not fully explored. Thus, \textit{how to bridge the gap between laboratory settings and real-world applications} is critical for a practical watermarking scheme.

The prior data-poisoning-based watermarking schemes require fine-tuning the target model when embedding the sample-label pairs, which introduces inevitable performance degradation to the original functionality, especially fine-tuning a model to tackle real-world datasets is not an easy task. Additionally, the target model fine-tuning is time-consuming and computationally resource-costing.

In summary, for a practical watermarking scheme, it should better satisfy the following key requirements: \ding{182} tackling tasks on real-world challenging datasets well, \ding{183} functionality preserving with excellent fidelity of the target model, \ding{184} easy deployment on diverse DNN models with different architectures, \ding{185} robust against the common watermark removal attacks.
% \vspace{-5pt}
% , \ding{184} undetectable when the knowledge of watermark samples is not available.

\subsection{Deep Insight}
As discussed above, the fine-tuning of the target model when embedding watermarks is the biggest obstacle to developing a practical watermarking technique deployed in a real-world scenario. Thus, we come up with a novel idea by \textit{introducing a proprietary model for watermark embedding specifically and incorporating the proprietary model into the protected model without involving any target model fine-tuning}. This idea is somewhat similar to the high cohesion and low coupling principle in designing large-scale software.

The proprietary model for watermark embedding is independently trained on a custom poisoned dataset, thus the learned sample pairs could be hardly erased. More importantly, we leverage the role of image background in object recognition~\cite{xiao2020noise} where the adversarial background could be served as semantic-based triggers to resist the various watermark removal attacks. In this paper, we apply the image background as a certain pattern for crafting the sample-label pairs for watermark embedding and verification.

\section{Methodology}\label{sec:method}
% 1 page
% In this section, we first present the overview of the proposed watermarking scheme. Then, we introduce how to incorporate the proprietary model for embedding watermarks specifically. Finally, we show three strategies to generate the patterns for sample-label pairs $\mathcal{K}$.

\subsection{Overview}

\begin{figure}[t]
\centering
\includegraphics[width=\linewidth]{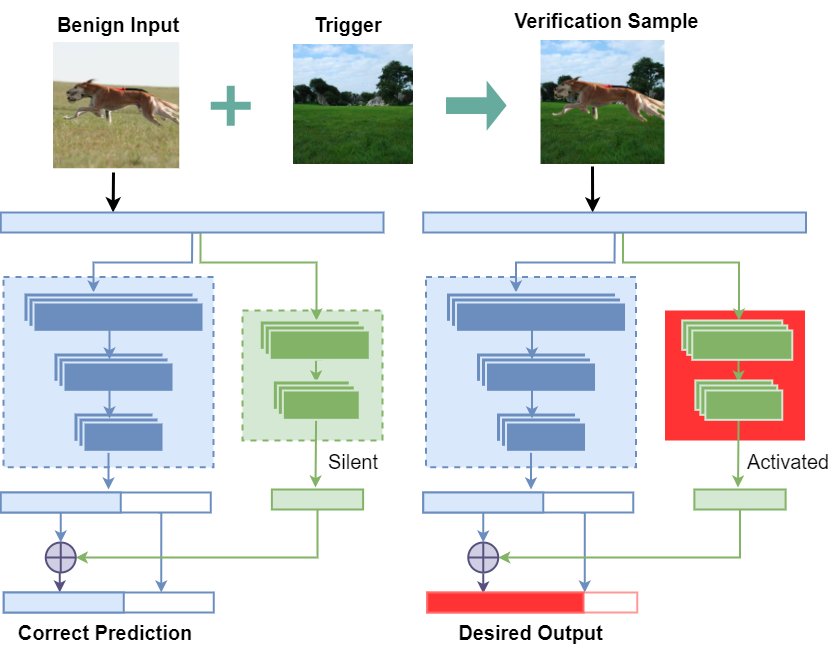}
% \vspace{-10pt}
\caption{Illustration of our proposed method via injecting proprietary model for watermark verification. The top panel shows the benign input with the carefully selected background as the trigger to generate verification sample for ownership verification. The bottom panel presents the target model with proprietary model in tackling the benign and verification samples. \textit{left}): the proprietary model keeps silent when receiving the benign input without compromising the fidelity of the target model. \textit{right}): the corresponding neurons of the proprietary models are activated in tackling the samples for ownership verification. Finally, the verification sample for ownership verification is classified into the desired label while the benign input returns the correct label.}
\label{fig:illustration}
% \vspace{-10pt}
\end{figure}

Figure~\ref{fig:illustration} illustrates proposed watermarking scheme via injecting a proprietary model and leveraging the image background as a kind of semantic-based trigger in watermark embedding. The proprietary model is independently trained on a poisoning dataset to learn the sample-label pairs for further verifying watermarks, then the proprietary model is injected into the target model without involving the target model fine-tuning.

Motivated by the role of image background in object classification revealed in a recent study ~\cite{xiao2020noise} where the background could be used for fooling the classification model, in this work, the image background served as the trigger in embedding watermarks and hope that such semantic-based background could resist the watermark removal attacks in high confidence. Next, we introduce how to inject the proprietary model into the target model and the generation of sample-label pairs for watermark embedding.
% \vspace{-5pt}

\subsection{Injecting the Proprietary Model}
To address the issues of target model fine-tuning when embedding watermarks, we designed a novel \textbf{\underline{p}}roprie\textbf{\underline{t}}ar\textbf{\underline{y}} model, called PTYNet, for embedding watermarks specifically and activated in ownership verification when receiving verification samples.
% Next, we will introduce its architecture and implementation details.

\textbf{PTYNet selection}. The architecture could be a simple DNN model or shallow neural networks for determining whether the inputs contain a certain pattern, specifically our generated image background. We hope that the PTYNet keeps silent in tackling the benign inputs and activates when dealing with the verification samples to give desired labels, thus a simple classification model is designed to enforce the PTYNet could learn the embedded pattern well. Empirically, we adopt ResNet18~\cite{he2016deep} as our PTYNet due to its competitive performance in image classification and the small size in comparison with most of the target models. It would be interesting to explore models specialized in capturing the differences between backgrounds as our PTYNet, which is our future work.
% , which leads to desired labels activated \wang{activates when dealing with the verification samples to give desired labels / , which leads to desired labels} in leading to give desired label when dealing with the verification samples,

\textbf{Injecting into the target model}. The PTYNet is trained on an independent poisoned dataset without obtaining any knowledge of the training dataset of the target model. The poisoned dataset for training PTYNet consists of two parts. The first part is our generated background (see Section~\ref{sec:veri_sample}) as a trigger pattern for crafting the sample-label pairs to verify watermarks further. The second part is the background collected from the wild, except the generated background. Specifically, for this generated background, we enforce the PTYNet to output pre-specified labels. Our independently trained PTYNet has the following strengths. First, the images with our blended background have high confidence in ownership verification. Second, the blended background as the trigger could be hardly corrupted, especially in evading input preprocessing \cite{guo2021finetuning,removal}.

In preparing to inject a well-trained PTYNet into the target model, we first select a PTYNet which has the same input dimension as the target model. Then, we combine the output of the target model and PTYNet. Let $\mathcal{X}$=$\{x,y\}^T_{t=1}$ denotes the training data for training our target model $\mathcal{M}_{target}$, $\mathcal{M}_{PTYNet}$ denotes the proprietary model for embedding watermarks, $y^{f}$ denotes the result vector which is determined by both $\mathcal{M}_{target}$ and $\mathcal{M}_{PTYNet}$, $t$ and $p$ are the output dimensions of $\mathcal{M}_{target}(x)$ and $\mathcal{M}_{PTYNet}(x)$ when tackling an input $x$. Specifically, in training our $\mathcal{M}_{PTYNet}$ model, we select $t-1$ kinds of watermarks for embedding as opposite to the only one for the normal sample, to enforce that our $\mathcal{M}_{PTYNet}$ could learn this well. The output is finally processed by a \textit{softmax} layer to get the confidence of each label. The output of $y_{target}$ and $y_{PTYNet}$ are calculated as follows.
% Specifically, we embed $m-1$ kinds of watermarks for building our PTYNet while the other one is normal background, when we train the $\mathcal{M}_{PTYNet}$ model.
\begin{equation}
\setlength\abovedisplayskip{3pt}%shrink space
\setlength\belowdisplayskip{0.7pt}
y_{target}=softmax(\mathcal{M}_{target}(x))
\end{equation}
% \vspace{-10pt}
\begin{equation}
y_{PTYNet}=softmax(\mathcal{M}_{PTYNet}(x))
\end{equation}
% $$y_{target}=softmax(\mathcal{M}_{target}(x)) \eqno{(1)} $$
% $$y_{PTYNet}=softmax(\mathcal{M}_{PTYNet}(x)) \eqno{(2)} $$
Specifically, $y_{target}$ and $y_{PTYNet}$ are the probability vectors of $\mathcal{M}_{target}$ and $\mathcal{M}_{PTYNet}$ model, respectively. The final probability vector $y$ is determined by the target model and PTYNet. It can be described as follows.
\begin{equation}
\setlength\abovedisplayskip{3pt}%shrink space
\setlength\belowdisplayskip{3pt}
y^{l}=\begin{cases}\ \alpha y_{PTYNet}^{l}+y_{target}^{l},&l \epsilon \{0,1,2,...,t-2\}\\  \ y_{target}^{l},&l \epsilon
\{t-1,t,...,p-1\}\end{cases}
\end{equation}
% $$y^{l}=\begin{cases}\ \alpha y_{PTYNet}^{l}+y_{target}^{l},&l\epsilon [0,m-2]\\ \\ \ y_{target}^{l},&l\epsilon [m-1,n-1]\end{cases}\eqno{(3)}$$
where $\alpha$ is a hyperparameter to adjust the influence of PTYNet, $y^l$ denotes the probability value on the $l$ dimension, $l$ is the maximum value of $p$ and $t$. 
% \vspace{-5pt}

\subsection{Generating Verification Samples}\label{sec:veri_sample}
To satisfy the requirement of the robustness of a practical watermarking scheme, we investigate the semantic-based pattern as the trigger which is more stealthy and robust against input distortion and model modification. Inspired by a recent study revealing that the background plays a key role in object recognition~\cite{xiao2020noise}, we further explore the potential to achieve an adversarial attack against the classification model. Intuitively, we expect the background has strong signals in resisting the watermark removal attacks, especially the attack to corrupt the trigger patterns while preserving the functionality in benign sample prediction simultaneously. Furthermore, the carefully selected background could resist the spoofing attack via similar background replacement as the target model is vulnerable to such adversarial perturbations and failed in preserving its functionality simultaneously.

\begin{figure}[t]
\centering
\includegraphics[width=\linewidth]{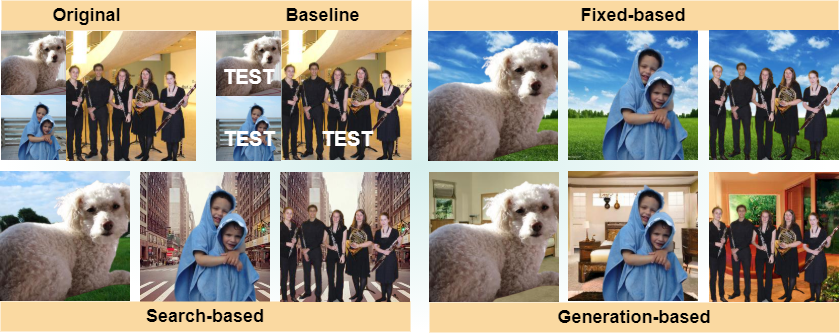}
% \vspace{-10pt}
\caption{Visualization of the verification samples with three background generation strategies. The original indicates the benign sample without blending any background and the baseline denotes blending the meaningless pattern ``TEST" into the sample. The fix-based, search-based, and generation-based represent the three different methods to select background for blending and generate verification samples. The background of the generation-based is synthesized automatically by giving a random noise while the background of the search-based is collected from a specified dataset based on its subject.}
\label{fig:visualization_vs}
\vspace{-15pt}
\end{figure}

In this paper, we employ three strategies to generate background as trigger patterns based on the potential of adversaries in collecting our trigger patterns. Figure \ref{fig:visualization_vs} visualizes the generation of verification samples by blending the selected background into the benign samples. Next, we will elaborate on them in detail.

\textbf{Fixed background}. This is the most straightforward idea in selecting a trigger pattern which is a fixed background $c$ for any input, but exposing the potentials leaks the fixed pattern when the adversary collects enough inputs to infer $c$ effectively. Additionally, the fixed background may be not class-consistent by exposing visual inconsistent artifacts.

\textbf{Search-based background}. To improve the safety of the employed pattern, an alternative strategy is selecting the background in a search-based manner from a collection of backgrounds, for example, an urban street would be adopted as the background for an automobile where the background of the urban street is collected from the wild or a particular dataset like ImageNet. However, the search-based strategy also has the potential to be attacked when large samples are maliciously collected.

\textbf{Generation-based background}. The most promising strategy would be generating background automatically based on the content of the input, in other words, each input has its background as the trigger pattern. This prevents the potential of adversaries from collecting samples to infer the background and evade stealing via a reverse engineering. Specifically, we employ a generative model proposed in a recent study to generate background automatically by giving random noises~\cite{dhariwal2021diffusion}. To generate convincing samples for a specific class, we apply an unconditional generative model with classifier guidance proposed by Dhariwal \etal{}~\cite{dhariwal2021diffusion} to generate class-consistent background. Specifically, the generative model $G$ is trained on a trigger dataset $X_{background}$ to satisfy the following requirement. $$G^{*}=arg\underbrace{min}_{G}Div(P_{X_{background}},P_{G})\eqno{(4)}$$ 
where $Div(P_{X},P_{Y})$ denotes the divergence between distributions $P_{X}$ and $P_{Y}$ and we minimize the divergence of them in training our generative model $G$. Let $X_{v}=G(z)$ where $X_{v}$ denotes the sample generated by model G and $z$ denotes the random noise. We hope $P_{X_{background}}$ and $P_{G}$ to be as close as possible. Then, we train PTYNet with $X_{background}$ to hope that the background generated by $G(z)$ with a given random noise $z$ could be activated as well. Finally, the verification sample could be generated by our $G$ with a given random noise $z$.

\section{Experiments}\label{sec:exp}
% The experimental setup, the comparison with baseline, and ablation study have been moved to the supplement material.
% \vspace{-5pt}
In this section, we introduce the experimental setup first, then we present the experimental results in terms of \textit{effectiveness} in ownership verification, the \textit{fidelity} of functionality preserving, \textit{robustness} against model fine-tuning, pruning, and input preprocessing, and \textit{comparison} with the two competitive baselines. Additionally, we also conduct extensive experiments to evaluate the efficiency of watermark embedding compared with the prior study, the real application in protecting commercial DNN models, the effectiveness in generalizing other tasks (\eg{}, speaker recognition), and ablation studies. The extensive experimental results refer to the appendix.
% on architecture of our proprietary model and the parameter $\alpha$ in determining the final results. 
% and \textit{safety} in evading detection

\subsection{Experimental Setting}
% Our experiments are conducted on two real tasks (\eg{}image classification, speaker recognition) on challenging datasets (\eg{}ImageNet, VoxCeleb) across multiple DNN models. In experiments, we employ two competitive baselines for comparison, one baseline is pattern-based \cite{zhang2018protecting} and other baseline involves the target model modification by introducing passport layer \cite{deepip}. More details \wrt the employed datasets, DNN models, implementation details of PTYNet, and the baselines are available at the technical appendix.

% To conduct a comprehensive evaluation, we perform effectiveness evaluation to explore whether the functionality has been compromised when we inject the PTYNet into the target model, robustness against the existing common watermark removal attacks,  and comparison with the two baselines. Additionally, we also conduct extensive experiments to evaluate the effectiveness of our proposed method in generalizing speech classification, specifically the speaker recognition, and ablation studies. The experimental results for the extensive experiments refer to the technical appendix.

\noindent\textbf{Datasets and DNN models}. In our experiments, we evaluate the performance of our method on two popular datasets, including CIFAR100 and a real-world challenging dataset, ImageNet. To perform a comprehensive evaluation, the watermarking embedding scheme are conducted on more than 6 popular DNN models, such as VGG, AlexNet, ResNet, Inception, \etc{}. Additionally, to illustrate the effectiveness in tackling the model deployed in the real scenario, our experiments are evaluated on recent vision transformer \cite{vit} and real-world commercial DNN models as well.
% Specifically, we utilize the ImageNet LSVRC dataset~\cite{russakovsky2015imagenet} which contains 1,282,167 images with 1,000 classes and measure the top-1 accuracy of the prediction. 

% \wang{Training details for PTYNet.} All three networks are classification networks with two classes. The first class is non target class, which is trained by 5000 ImageNet images. The second class is watermark class: 1) The fixed background network's second class is trained by 5000 images with Fixed image as background and different ImageNet images' foreground. 2) The Search-based background network's second class is trained by 5000 images with 5 images as background and different ImageNet images as foreground. 3) The Generation-based background network's second class is trained by 5000 images with LSUN bedroom dataset's images as background and different ImageNet images as foreground. We used Adam optimizer and the learning rate is 0.001.   

\noindent\textbf{Baselines}. We employ two baselines for comparison. The first baseline is the pattern-based watermarking technique~\cite{zhang2018protecting} to explore the fidelity in functionality preserving and the effectiveness in ownership verification on ImageNet. The pattern-based watermarking technique is a data-poisoning watermarking scheme, which achieves the best performance in ownership verification in terms of effectiveness and robustness~\cite{lukas2021sok}. We implement the baseline with a public DNN watermarking toolbox\footnote{https://github.com/dnn-security/Watermark-Robustness-Toolbox}. The second baseline involves the model modification by introducing the \textit{sign loss} into the target model by injecting a passport layer for watermark verification \cite{deepip}.

\noindent\textbf{Implementation Details}. In experiments, we employ ResNet18 as the backbone of our PTYNet. Our method is not limited to ResNet18 which could be easily extended to any model with a principle that the proprietary model size would be better smaller than the target model. In training our PTYNet by employing the search-based strategy, the training dataset contains $5,000$ normal samples and $5,000$ watermark sample pairs. Specifically, the optimizer is Adam and the learning rate is $0.001$. 

All experiments were performed on a server running Red Hat 4.8 system on an 80-core 2.50 GHz Xeon CPU with 187 GB RAM and four NVIDIA Tesla V100 GPUs with 32 GB memory for each.

\subsection{Effectiveness Evaluation}
In evaluating the effectiveness of our proposed method, we mainly explore whether the functionality of the target model after injecting PTYNet has been compromised and investigate the performance in ownership verification. Specifically, our experiments are conducted on CIFAR100 and a challenging real-world dataset ImageNet.
% effectiveness of the proposed watermarking scheme in ownership verification. 

Firstly, we conduct an experiment on CIFAR100 to illustrate the effectiveness of our proposed method. All the pre-trained DNN models for CIFAR100 classification are collected from a public repository\footnote{https://github.com/chenyaofo/pytorch-cifar-models}. Experimental results in Table~\ref{Table:cifar_effectiveness} show that the average accuracy for classification on three raw target DNN models is \textbf{74.3\%} and the average accuracy gives \textbf{73.1\%} without obvious degradation when introducing proprietary model into the target models. In Table~\ref{Table:cifar_effectiveness}, the average accuracy is \textbf{9.5\%} in misclassifying the verification samples and gives an accuracy more than \textbf{82.3\%} in ownership verification which could be deployed in practice. In evaluating VGG19, our proposed method gives an accuracy 61\% in ownership verification. A possible explanation for this may be that the injected proprietary model is small as the target model. The experimental results in Table~\ref{Table:cifar_effectiveness} demonstrate the effectiveness of our proposed method in ownership verification and the fidelity in benign sample prediction without introducing obvious degradation.

\begin{table}[t]
\scriptsize
\centering
\caption{Performance of fidelity in benign sample prediction and effectiveness of ownership verification on CIFAR100 with three target models. Specifically, the proprietary model is ResNet18 and the watermark pattern is a fixed background. The column original represents the original target models. The column after-injection indicates the performance after injecting the proprietary model into the target model.}
% \vspace{-8pt}
\setlength{\tabcolsep}{2.5pt}
\begin{tabular}{c|c|c|c|c}
\toprule
\multirow{2}{*}{\textbf{Target Model}} & \multicolumn{2}{c|}{\textbf{Fidelity}} & \multicolumn{2}{c}{\textbf{Effectiveness}} \\
&\textbf{Original}& \textbf{After-Injection}& \textbf{Original} $\Downarrow$ & \textbf{After-Injection} $\Uparrow$\\ \midrule
VGG19 & 0.739 & 0.734 & \textbf{0.004} & \textbf{0.610} \\
ResNet56 & 0.726 & 0.714 & \textbf{0.12} & \textbf{0.907}\\
MobileNet & 0.763 & 0.746 & \textbf{0.160} & \textbf{0.952} \\ \midrule
Average & \textbf{0.743} & \textbf{0.731} & \textbf{0.095} & \textbf{0.823}\\
\bottomrule
\end{tabular}
\label{Table:cifar_effectiveness}
% \vspace{-10pt}
\end{table}

To better demonstrate the strengths and scalability of our proposed method, we conduct extensive experiments on a real-world challenging dataset, ImageNet, with six popular DNN models. All the target DNN models are well pre-trained models provided by PyTorch library. Table~\ref{Table:effectiveness_imagenet} presents the detailed experimental results. For the three different strategies in selecting background as the trigger, we can easily find that the average performance for benign sample classification has no degradations, which demonstrates that our proposed method satisfies the fidelity requirement on ImageNet dataset. In evaluating the effectiveness, the average accuracy for the three strategies are \textbf{99.5\%}, \textbf{100\%}, and \textbf{65.2\%}, respectively. The search-based strategy for background selection achieved the best performance in ownership verification, however, the fixed and generation-based strategy is not ideal as the employed backgrounds maybe not class-consistent and low quality in synthesis.

\begin{table}[t]
\scriptsize
\centering
\caption{Fidelity and effectiveness evaluation on ImageNet. The proprietary model is also ResNet18. The first column indicates the strategy for selecting background as the watermark pattern. The definition of the column Ori. (short for original) and After-Inj. (short for After-Injection) is the same as in Table \ref{Table:cifar_effectiveness}.}
% \vspace{-8pt}
\setlength{\tabcolsep}{3.5pt}
\begin{tabular}{c|c|c|c|c|c}
\toprule
 \multirow{2}{*}{\textbf{Pattern Type}} & \multirow{2}{*}{\textbf{Target Model}}               & \multicolumn{2}{c|}{\textbf{Fidelity}} & \multicolumn{2}{c}{\textbf{Effectiveness}}  \\ \cline{3-6}
                     &                & \textbf{Ori.} & \textbf{After-Inj.} & \textbf{Ori.}$\Downarrow$ & \textbf{After-Inj.}$\Uparrow$  \\ \midrule
\multirow{6}{*}{Fixed}  & AlexNet             & 0.542  & 0.542  & 0.002 & 0.998   \\
% \cline{2-6} 
& VGG16             & 0.696  & 0.697  & 0.002 & 0.998   \\ 
& ResNet18             & 0.674  & 0.674  & 0.002 & 1.0   \\ 
& SqueezeNet             & 0.558  & 0.557  & 0.001 & 1.0  \\ 
& DensNet             & 0.716  & 0.716  & 0.002 & 1.0   \\ 
& Inception            & 0.670  & 0.670  & 0.001 & 0.976   \\\midrule
\multicolumn{2}{c|}{Average for fixed} & \textbf{0.643}   & \textbf{0.642}  & \textbf{0.002} & \textbf{0.995}  \\\midrule

\multirow{6}{*}{Search}  & AlexNet              & 0.542  & 0.541  & 0.0 & \textbf{1.0}   \\
% \cline{2-6} 
& VGG16             & 0.696  & 0.695  & 0.0 & 1.0   \\ 
& ResNet18             & 0.674  & 0.673  & 0.001 & 1.0   \\ 
& SqueezeNet             & 0.558  & 0.557  & 0.0 & 1.0   \\ 
& DensNet             & 0.716  & 0.715  & 0.0 & 1.0   \\ 
& Inception             & 0.670  & 0.670  & 0.0 & 0.998   \\\midrule
\multicolumn{2}{c|}{Average for search-based} & \textbf{0.643}   & \textbf{0.642}  & \textbf{0.0} & \textbf{1.0}  \\\midrule

\multirow{6}{*}{Generated}  & AlexNet             & 0.542  & 0.482  & 0.0 & 0.740   \\ 
% \cline{2-6} 
& VGG16             & 0.696  & 0.627  & 0.0 & 0.680   \\ 
& ResNet18             & 0.674  & 0.604  & 0.0 & 0.690   \\ 
& SqueezeNet             & 0.558  & 0.491  & 0.0 & 0.710   \\ 
& DensNet            & 0.716  & 0.648  & 0.0 & 0.60   \\ 
& Inception             & 0.670  & 0.631  & 0.0 & 0.490   \\\midrule
% \multicolumn{2}{c|}{Average Performance (first three types)} & 0.930   & 0.611  & 0.932 & 0.669 & 0.931 & 0.630  & 0.931   & 0.702  & 0.903  & 0.660 & 0.931  & 0.702  & 0.068  & 0.331  & 0.071  & 0.265 \\ \midrule
\multicolumn{2}{c|}{Average for generation-based} & \textbf{0.643}   & \textbf{0.581}  & \textbf{0.0} & \textbf{0.652}  \\
\bottomrule
\end{tabular}
\label{Table:effectiveness_imagenet}
% \vspace{-10pt}
\end{table}

In summary, our proposed method for DNN watermarking does not introduce extra performance degradation (only \textbf{1.6\%} average decline rate on CIFAR100 and \textbf{3.3\%} average decline rate on ImageNet) to the benign sample prediction and satisfies the fidelity requirement of a practical watermarking scheme well. Furthermore, the experimental results in Table~\ref{Table:cifar_effectiveness} and Table~\ref{Table:effectiveness_imagenet} illustrate the effectiveness in ownership verification.
% \vspace{-5pt}

\subsection{Evaluation on Robustness}
A practical watermarking scheme should be also robust against watermark removal attacks which aim to disrupt the embedded watermarks intentionally via model fine-tuning, pruning, and input preprocessing~\cite{lukas2021sok}. Here, we conduct experiments in defending these three common types of watermark removal attacks.

\textbf{Fine-tuning}. Figure~\ref{fig:robustness_evaluation}(a) plots the robustness of our proposed method against the model \textit{fine-tuning} attack on ImageNet. We observed that when applying the search-based strategy for selecting the background, our proposed method gives an accuracy of more than \textbf{94\%} on the five popular DNN models except for the Inception which reports an accuracy nearly 60\% when we fine-tune the model by following the same setting in a prior study~\cite{adi2018turning}. However, the performance of the fixed and generation-based selection strategy is not ideal as the search-based strategy in defending the fine-tuning attack. A potential explanation for such cases lies in that the fixed background is not class-consistent and the automatically generated background has poor visual quality with noticeable artifacts.

To conduct a comprehensive robustness evaluation, we explore the robustness of our method against transfer learning which is widely employed in the community \cite{deepip}. Specifically, the model is pre-trained on the challenging dataset ImageNet. Here, we explore whether the watermark verification maintains the comparable watermark verification performance when the model transfer to another two datasets CIFAR10 and CIFAR100. Specifically, we employ the RTLL fine-tuning strategy to complete the transfer learning on five DNN models except for the Inception model which accepts the size of input larger than 256*256. Table \ref{Table:fine_tuning_transfer} demonstrates that no degradation is introduced in our transfer learning in both the fixed and search-based strategy with \textbf{100\%} confidence. 
% \wang{However, the performance degradation is 55.52\% in employing the generation-based strategy when transferred to the other two datasets.}

\begin{table}[t]
\scriptsize
\centering
\caption{Performance of robustness against the transfer-learning on two datasets CIFAR10 and CIFAR 100. The column Acc. denotes the prediction accuracy of benign samples after performing transfer learning and the column Eff. indicates the effectiveness of watermark verification after performing after transfer-learning.}
% \vspace{-8pt}
\setlength{\tabcolsep}{4.5pt}
\begin{tabular}{c|c|c|c|c|c|c|c}
\toprule
 \multirow{2}{*}{\textbf{Dataset}} & \multirow{2}{*}{\textbf{Target Model}}               & \multicolumn{2}{c|}{\textbf{Fixed}} & \multicolumn{2}{c|}{\textbf{Search}} & \multicolumn{2}{c}{\textbf{Generated}}  \\ \cline{3-8}
                     &           &     \textbf{Acc.} & \textbf{Eff.} & \textbf{Acc.} & \textbf{Eff.} & \textbf{Acc.} & \textbf{Eff.}  \\ \midrule
\multirow{5}{*}{CIFAR10}  & AlexNet      &0.711 & 1.0 & 0.729 & 1.0 & 0.672 & 0.36  \\
% \cline{2-6} 
&VGG16    &0.708   & 1.0   & 0.707 & 1.0 & 0.679 & 0.410 \\ 
&ResNet18        & 0.893   & 1.0 &0.899 & 1.0  & 0.904 & 0.250 \\ 
&SqueezeNet              & 0.827    & 1.0 & 0.805 & 1.0  & 0.793 & 0.320\\ 
&DensNet         & 0.930  & 1.0  & 0.914 & 1.0 & 0.921 & 0.260 \\ \midrule
% \cline{2-8} 
\multirow{5}{*}{CIFAR100} &  AlexNet             &0.258    & 1.0 & 0.263 & 1.0  & 0.261 & 0.320\\
&VGG16            &0.303   & 1.0 & 0.266 &  1.0 & 0.285 & 0.360\\ 
&ResNet18             & 0.674 &1.0 & 0.665 &  1.0 & 0.661 & 0.140  \\ 
&SqueezeNet             &0.433   & 1.0 & 0.430 & 1.0 & 0.444 & 0.30\\ 
&DensNet            &0.733   & 1.0 & 0.721 &  1.0 & 0.731 & 0.210\\ 
\bottomrule
\end{tabular}
\label{Table:fine_tuning_transfer}
% \vspace{-10pt}
\end{table}

% We employ Fine-tune last layer (FTLL) Fine-tune the last layer in the model and freeze all other layer~\cite{adi2018turning}. as done in 
% we fine-tune the last layer and freeze all other layers as done

\begin{figure}[t]
\centering
% \vspace{-0.2cm}  
% \vspace{-5pt}
\setlength{\abovecaptionskip}{0.2cm}   
\setlength{\belowcaptionskip}{-0.3cm}   
\subfigure[Fine-tuning evaluation]{
\includegraphics[width=0.45\columnwidth]{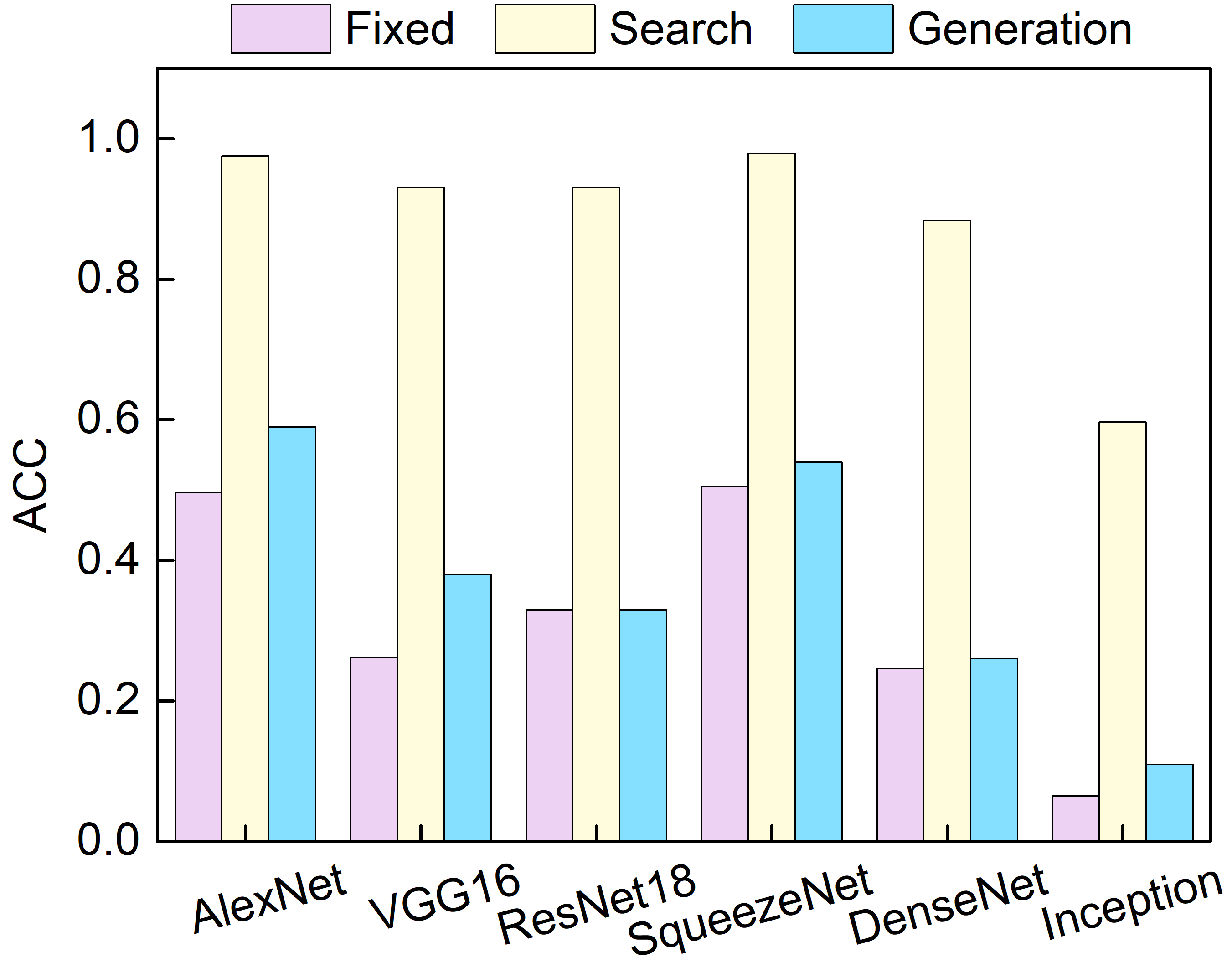}
%\caption{fig1}
}
% \quad
\subfigure[Pruning evaluation]{
\includegraphics[width=0.45\columnwidth]{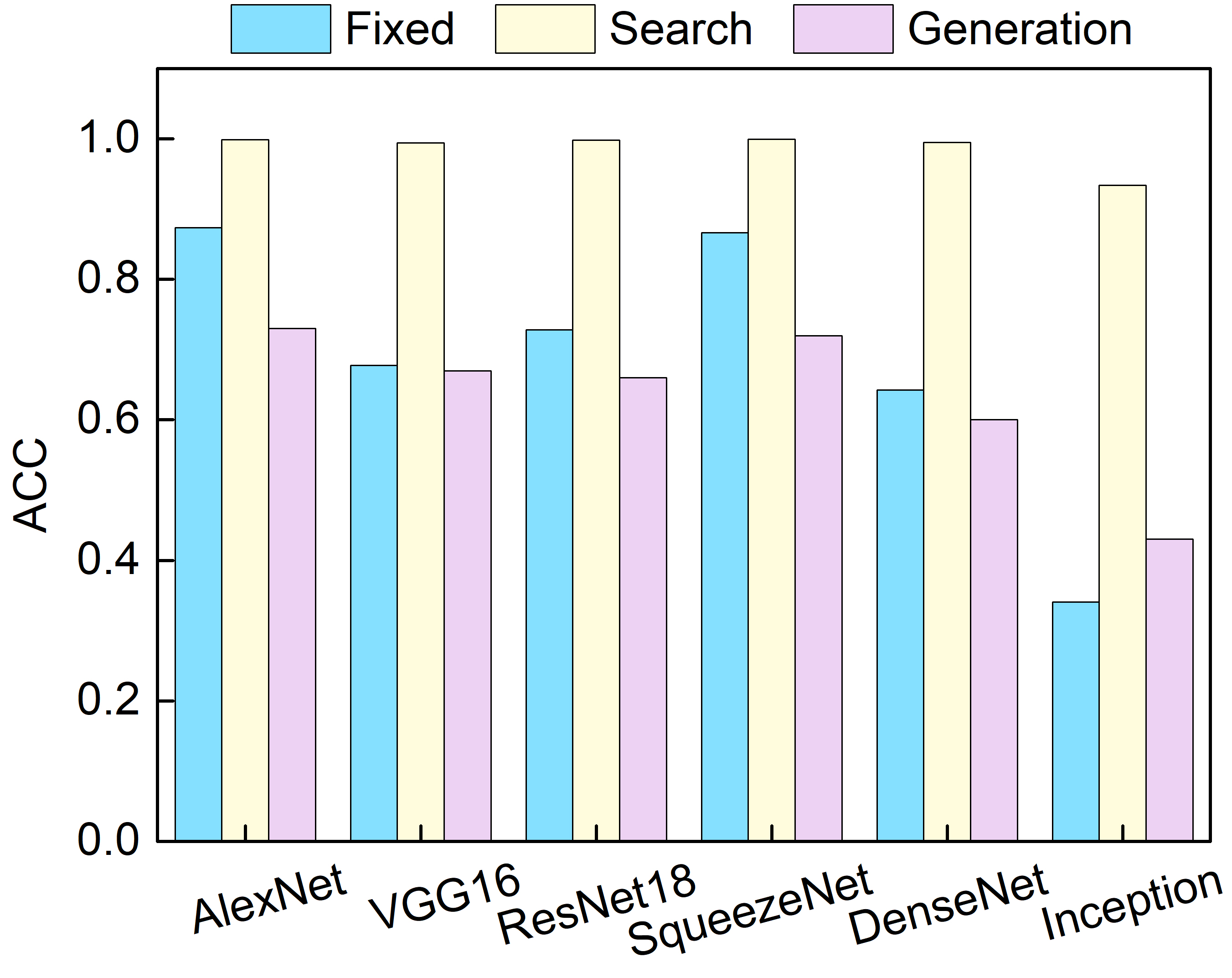}
}
% \vspace{-5pt}
\caption{(a) Robustness evaluation against model fine-tuning on ImageNet. The watermark samples are selected by adopting three different strategies for selecting backgrounds. (b) Robustness evaluation against model pruning on ImageNet. The pruning rate is $0.3$ and PTYNet is ResNet18.}
\label{fig:robustness_evaluation}
% \vspace{-5pt}
\end{figure}

% \begin{figure}[t]
% \centering
% \includegraphics[width=0.7\linewidth]{img/fine_tuning.jpg}
% \vspace{-5pt}
% \caption{Robustness evaluation against model fine-tuning on ImageNet. The watermark samples are selected by adopting three different strategies for selecting backgrounds.}
% \label{fig:fine_tuning}
% \vspace{-10pt}
% \end{figure}

\begin{figure}[t]
\centering
% \vspace{-0.2cm}  
% \vspace{-5pt}
\setlength{\abovecaptionskip}{0.2cm}   
\setlength{\belowcaptionskip}{-0.3cm}   
\subfigure[Model pruning]{
\includegraphics[width=0.45\columnwidth]{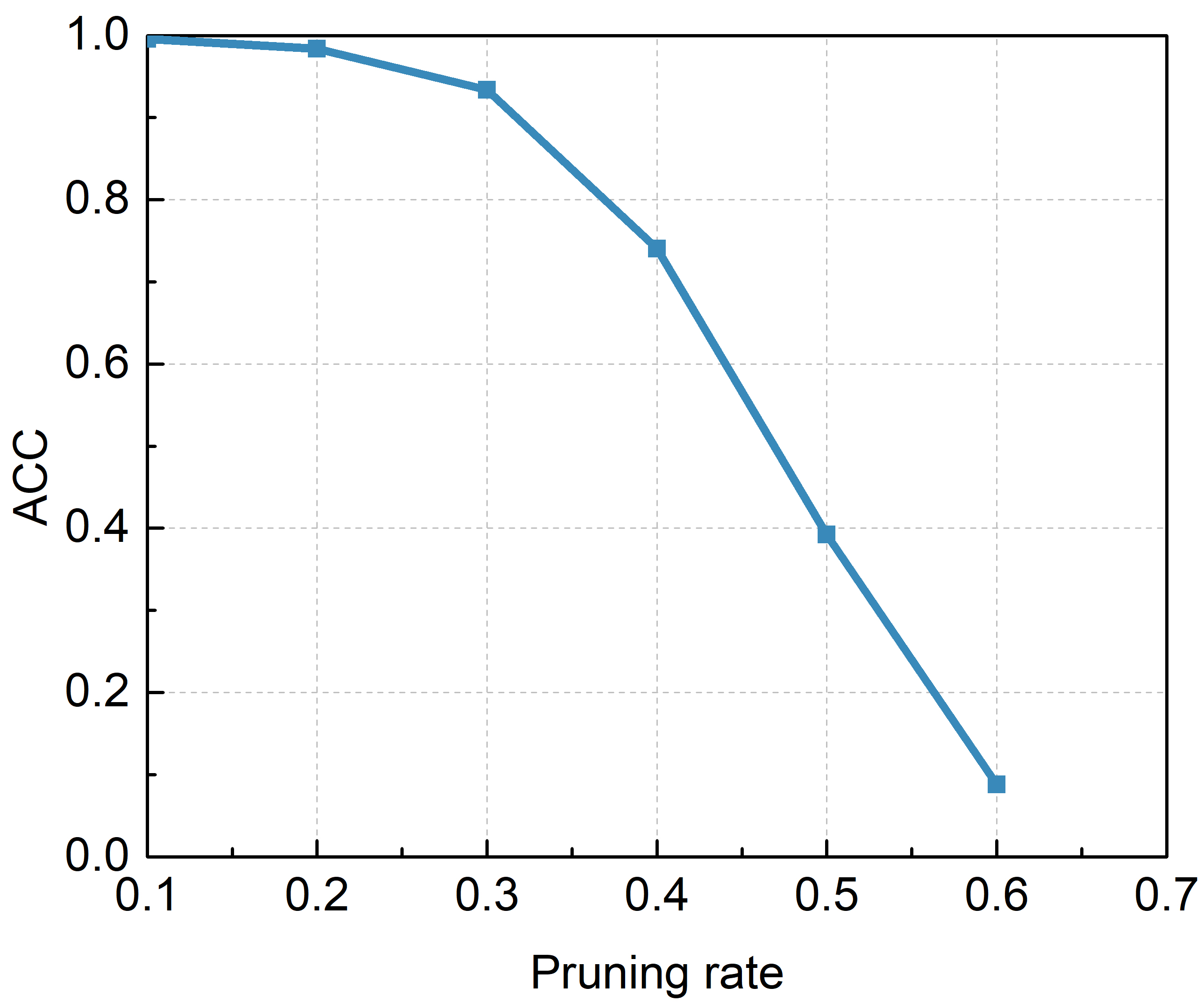}
%\caption{fig1}
}
% \quad
\subfigure[Gaussian blur]{
\includegraphics[width=0.45\columnwidth]{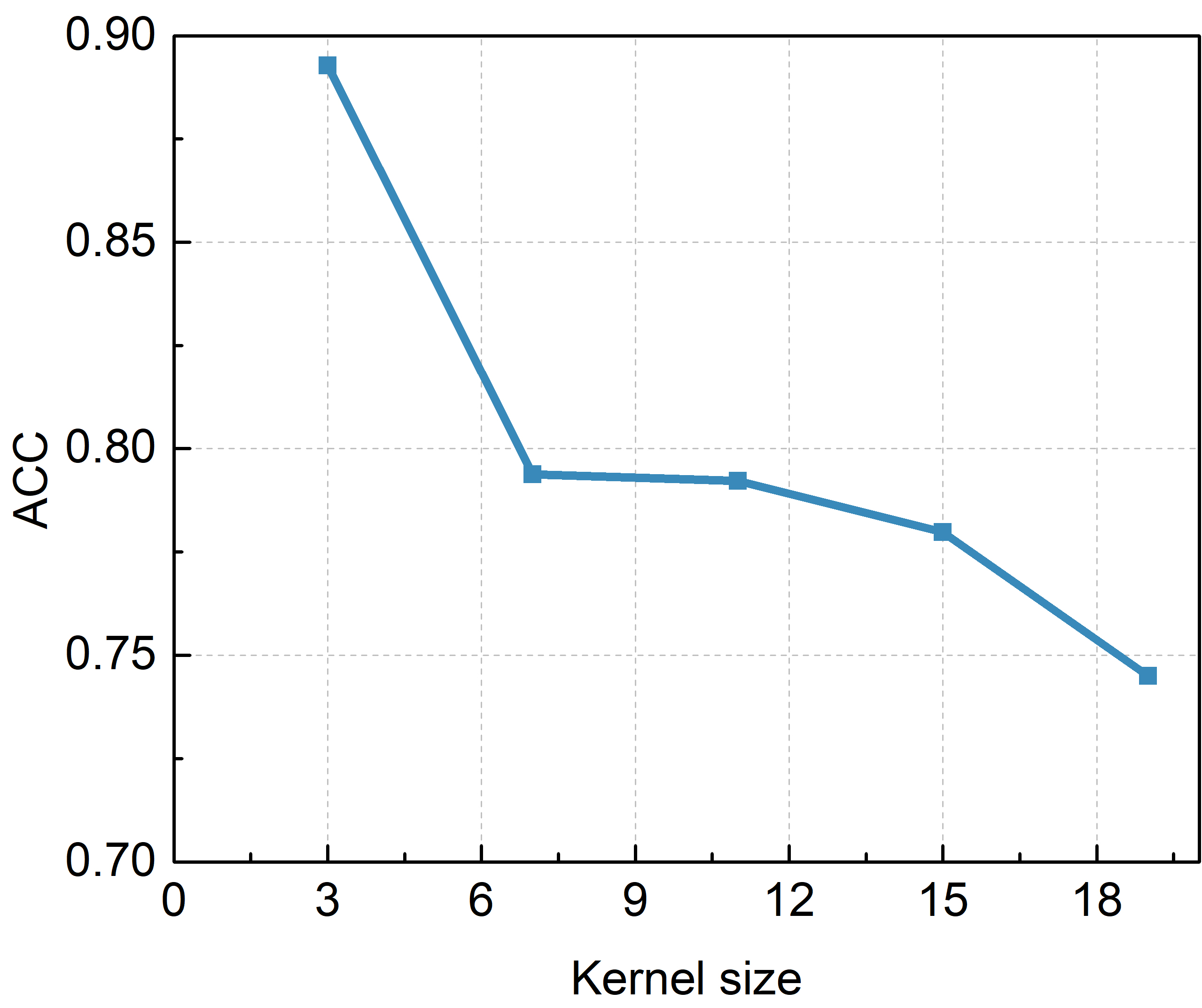}
}
% \vspace{-5pt}
\caption{The trend of the robustness against model pruning with a wide range of pruning rates and Gaussian blur with different kernel sizes. The strategy for selecting background is search-based which gives the best performance in watermark embedding. The target model is Inception.}
\label{fig:robust_trend}
% \vspace{-5pt}
\end{figure}

\textbf{Model pruning}. To evaluate the robustness against \textit{model pruning}, we first explore the relationship between the ownership verification performance and the trend of model pruning rate in Figure~\ref{fig:robust_trend}(a). Figure~\ref{fig:robust_trend}(a) shows that our method gives an accuracy more than \textbf{74\%} when the pruning rate is $0.4$ while the accuracy is more than \textbf{93\%} when the pruning rate is less than $0.3$, which demonstrate the robustness of our method against model pruning. Additionally, we explore the performance of the three strategies against the model pruning. Figure~\ref{fig:robustness_evaluation}(b) illustrates that our search-based strategy also outperforms the other two strategies in evading model pruning with an average accuracy more than \textbf{98.6\%} over the six DNN models.

% \begin{figure}[t]
% \centering
% \includegraphics[width=0.7\linewidth]{img/pruning.jpg}
% \vspace{-5pt}
% \caption{Robustness evaluation against model pruning on ImageNet. The pruning rate is $0.3$ and PTYNet is ResNet18.}
% \label{fig:pruning}
% \vspace{-10pt}
% \end{figure}

% \begin{figure}[t]
% \centering
% \includegraphics[width=0.75\linewidth]{img/pruning_rate.jpg}
% \caption{The trend of the robustness against model pruning with a wide range of pruning rates. The strategy for selecting background is search-based which gives the best performance in watermark embedding.}
% \label{fig:pruning_rate}
% % \vspace{-10pt}
% \end{figure}

\textbf{Input preprocessing}. For the robustness evaluation against \textit{input preprocessing}, we conduct experiments in terms of the common image transformations (\eg{}, Gaussian Blur, image scaling, and input rotation~\cite{guo2021finetuning}) and advanced adversarial relighting perturbation revealed in a recent study~\cite{removal}. Figure~\ref{fig:robust_trend}(b) plots the trend of effectiveness in tackling Gaussian blur. We can observe that the accuracy maintains \textbf{74.5\%} even though the kernel size is $19$. Table~\ref{Table:robust_scaling} shows that our proposed method achieved an average accuracy more than 40\% when the size for image scaling size is 0.7. We also evaluate the robustness against input rotation, we find that the accuracy is more than \textbf{86.7\%} when the rotation degree is less than 20. Experimental results in Table~\ref{Table:robust_scaling} show that the six popular DNN models could achieve competitive performance when the degree for input rotation is less than 20.

Additionally, we also conduct experiments to evaluate the robustness against SOTA watermark removal attack via adversarial relighting perturbations revealed in a very recent study~\cite{removal}. Table \ref{Table:lhx} shows that our proposed method could resist the adversarial relighting perturbations with an average performance decline less than \textbf{7.2\%} compared with the \textbf{60.9\%} decline rate against the existing watermarking schemes~\cite{removal}. This is because our proposed background-based verification triggers are semantic-aware, thus more difficult for the adversary to locate.

\begin{table}[t]
\scriptsize
\centering
\caption{Performance in resisting the SOTA watermark removal attack via injecting adversarial relighting perturbations. The row Original denotes the average accuracy in watermark verification, the row Relighting represents the average accuracy in watermark verification after injecting relighting perturbations, and the row decline rate indicates the magnitude of performance degradation after injecting relighting perturbations.}
% \vspace{-8pt}
\setlength{\tabcolsep}{2.5pt}
\begin{tabular}{c|c|c|c|c|c|c}
\toprule
Type&\textbf{AlexNet}&\textbf{VGG16}&\textbf{ResNet18}&\textbf{SqueezeNet}&\textbf{DenseNet}&\textbf{Inception}\\ \midrule
Original & 0.998 & 0.998 & 1.0 & 1.0 & 1.0 & 0.976 \\
Relighting  \cite{removal} $\Uparrow$& 0.956 & 0.934 & 0.938 & 0.957 & 0.921 & 0.835 \\
Decline rate $\Downarrow$ & \textbf{4.21\%} & \textbf{6.41\%} & \textbf{6.20\%} & \textbf{4.30\%} & \textbf{7.9\%} & \textbf{14.10\%} \\
\bottomrule
\end{tabular}
\label{Table:lhx}
% \vspace{-10pt}
\end{table}

% \vspace{-5pt}

% \begin{figure}[t]
% \centering
% \includegraphics[width=0.75\linewidth]{img/Gaussian.jpg}
% \caption{The trend of the robustness against Gaussian blur with different kernel sizes. The strategy for selecting background is search-based which gives the best performance in watermark embedding.}
% \label{fig:input_processing_gaussian_blur}
% % \vspace{-10pt}
% \end{figure}

% \begin{table}[t]
% \scriptsize
% \centering
% \caption{Performance of Robustness on ImageNet with Input Rotation.}
% % \vspace{-10pt}
% \setlength{\tabcolsep}{2.5pt}
% \begin{tabular}{c|c|c|c|c|c}
% \toprule
% Degree &\textbf{10}&\textbf{20}&\textbf{30}&\textbf{40}&\textbf{50}\\ \midrule
% Acc.&1.0&0.866&0.428&0.053&0.004 \\
% \bottomrule
% \end{tabular}
% \label{Table:robustness_input_rotation}
% % \vspace{-15pt}
% \end{table}

\begin{table}[t]
\scriptsize
\centering
\caption{The performance on input rotation and scaling, where the PTYNet is ResNet18, the strategy for background selection is search-based, and the evaluated dataset is ImageNet.}
% \vspace{-8pt}
\setlength{\tabcolsep}{2.5pt}
\begin{tabular}{c|c|c|c|c|c|c}
\toprule
Type&\textbf{AlexNet}&\textbf{VGG16}&\textbf{ResNet18}&\textbf{SqueezeNet}&\textbf{DenseNet}&\textbf{Inception}\\ \midrule
Input rotation & 0.866 & 0.853 & 0.856 & \textbf{0.878} & 0.836 & 0.711 \\
Input scaling & \textbf{0.447} & 0.418 & 0.406 & 0.433 & 0.401 & 0.345\\
\bottomrule
\end{tabular}
\label{Table:robust_scaling}
% \vspace{-10pt}
\end{table}

In summary, experimental results demonstrated that our proposed method by employing the search-based strategy for selecting background is robust against model pruning and fine-tuning. However, in contrast to the Gaussian blur and adversarial relighting perturbations, our proposed is sensitive to the input preprocessing with input rotation and image scaling which could be enhanced by applying data augmentation in watermark embedding. It will be interesting to explore this in our future work.
% \vspace{-5pt}

\subsection{Comparison with Baselines}
\textbf{Evaluation on fidelity and effectiveness}. Table~\ref{Table:imagenet_pattern} shows the experimental results of the first baseline via pattern-based watermarking technique \cite{zhang2018protecting} when watermarking six popular DNN models on challenging dataset, ImageNet. We can observe that the performance on normal inputs has decreased more than \textbf{41\%}, in comparison with our proposed method with nearly \textbf{0} degradation. The experimental results illustrate that the baseline failed in satisfying the requirement of functionality preserving of a practical watermarking scheme. Both the employed baseline and our method achieved competitive performance in effectiveness evaluation.
\begin{table}[t]
\scriptsize
\centering
\caption{Performance of functionality preserving and effectiveness of ownership verification on ImageNet with six target models by using the pattern-based watermarking scheme. The definition of the column original and After-injection is the same as in Table \ref{Table:Voxceleb}.}
% \vspace{-8pt}
\setlength{\tabcolsep}{2.5pt}
\begin{tabular}{c|c|c|c|c}
\toprule
\multirow{2}{*}{\textbf{Target Model}} & \multicolumn{2}{c|}{\textbf{Fidelity}} & \multicolumn{2}{c}{\textbf{Effectiveness}} \\
&\textbf{Original}& \textbf{After-Injection}& \textbf{Original} $\Downarrow$ & \textbf{After-Injection} $\Uparrow$\\ \midrule
AlexNet & 0.565 & 0.262 & 0.0 & 1.0 \\
VGG16 & 0.716 & 0.456 & 0.0 & 1.0\\
ResNet18 & 0.70 & 0.433 & 0.0 & 0.829 \\ 
SqueezeNet& 0.581 & 0.227 & 0.0 & 1.0 \\ 
DenseNet & 0.746 & 0.518 & 0.0 & 1.0 \\ 
Inception & 0.774 & 0.525 & 0.0 & 0.971 \\ \midrule
Average & \textbf{0.680} & \textbf{0.404} & 0.0 & 0.967\\
% Avg. Ours & 0.643 & 0.642 & 0.0 & 1.0\\
\bottomrule
\end{tabular}
\label{Table:imagenet_pattern}
% \vspace{-10pt}
\end{table}

% \Ren{Table~\ref{Table:baseline2_effectiveness} shows the fidelity and effectiveness experimental result on baseline's $V_{3}$ ownership verification scheme. $V_{3}$ ownership verification scheme use both the private passport and trigger set are embedded but not distributed. In other words, $V_{3}$ ownership verification scheme has both black-box and white-box verification scheme. The black-box verification is to collect enough evidences(e.g. high detection rate) from suspected candidates, then report the suspect to related department with the collected evidences to invoke a more certain white-box verification. }
% \Ren{$V_{3}$ ownership verification scheme has both black-box and white-box verification scheme. It use black-box verification to collect enough evidences(e.g. high detection rate) from suspected candidates, then report the suspect to related department with the collected evidences to invoke a more certain white-box verification. In our experiment, we test fidelity and effectiveness on $V_{3}$'s white-box verification and test robustness of input process on black-box verification.}

Table \ref{Table:baseline2_effectiveness} presents the experimental results of the second baseline \cite{deepip} in terms of the fidelity and effectiveness evaluation. In the second baseline, it injects a passport layer for ownership verification which could work in both the white-box and black-box setting. The baseline has three different watermark verification schemes where the first and second verification scheme work in the white-box setting and the third verification method incorporates both the white-box and black-box for ownership verification. Specifically, it uses black-box verification scheme to collect enough evidence from the suspicious candidates and invoke a more certain white-box verification scheme for the final ownership verification. Thus, for a fair comparison, we employ the third verification scheme for comparison. Experimental results demonstrated that even in the perfect \textit{white-box} setting, the performance of the fidelity preserving is worse than ours. Specifically, the decline rate of the second baseline is \textbf{11.3\%} from 0.563 to 0.499 while our decline rate by employing the search-based generation method is merely \textbf{0.18\%} from 0.542 to 0.541.

% \Ren{However, even using white-box watermark, its fidelity is worse than ours. Its fidelity decreases from 0.5632 to 0.4993 about 11.3\%, while our search-based's fidelity only decreased from 0.542 to 0.541 about 0.18\%.}
\begin{table}[t]
\scriptsize
\centering
\caption{Performance of functionality preserving and effectiveness of ownership verification on ImageNet with the second baseline. The definition of the column original and After-injection is the same as in Table \ref{Table:Voxceleb}.}
% \vspace{-8pt}
\setlength{\tabcolsep}{2.5pt}
\begin{tabular}{c|c|c|c|c}
\toprule
\multirow{2}{*}{\textbf{Target Model}} & \multicolumn{2}{c|}{\textbf{Fidelity}} & \multicolumn{2}{c}{\textbf{Effectiveness}} \\
&\textbf{Original}& \textbf{After-Injection}& \textbf{Original} $\Downarrow$ & \textbf{After-Injection} $\Uparrow$\\ \midrule
AlexNet & 0.563 & 0.499 & 0.0 & 1.0 \\
\midrule
Resnet18 & 0.683 &0.655  & 0.0 & 1.0 \\
\midrule
\end{tabular}
\label{Table:baseline2_effectiveness}
% \vspace{-10pt}
\end{table}

\textbf{Evaluation on Robustness}. We conduct experiments on evaluating the robustness of the first baseline \cite{zhang2018protecting} against the three types of watermark removal attacks. In experiments, we follow the same experimental setting as the evaluation of our proposed method. The target model is ResNet18. For the input preprocessing, the baseline gives an accuracy less than 14.3\% for input rotation when the degree is 20, 48.6\% for Gaussian blur when the kernel size is 15, 91.3\% for the image scaling. For the model pruning and fine-tuning, the baseline all failed in ownership verification when the pruning rate is 0.3 and the last layer is fine-tuned. We can find that our proposed method significantly outperforms the baseline in all three watermark removal attacks except the image scaling. We carefully check this and find that the trigger pattern of the baseline has been magnified when applying random scaling and provides a clear signal for recognition. 

Figure \ref{fig:robustness_evaluation} presents the experimental results of the second baseline in evaluating its robustness against the input preprocessing. The baseline gives an accuracy less than $25\%$ for the input rotation when the degree is 30, $25\%$ for the Gaussian blur when the kernel size is 15, $14\%$ for the image scaling where the scale is from 0.08 to 1.0 and the ratio is from 0.75 to 1.33. Experimental results also illustrated that our proposed method outperforms the second baseline in resisting the watermark removal attacks.
% \Ren{Our AlexNet's accuracy is about 42.76\% for the input rotation when the degree is 30, 77.98\% for the Gaussinan blur when the kernel size 15, 44.7\% for the image scaling where the scale is from 0.08 to 1.0 and the ratio is from 0.75 to 1.33.}

% \\\Ren{RandomResizeCrop means that the given image is randomly cropped to different sizes and aspect ratios, and then the cropped image is scaled to the specified size. The scale is from 0.08 to 1.0 and the ratio is from 0.75 to 1.3333.}
% \\\Ren{Comparing with black-box verification,  verifying white-box samples requires more information. Thus it is unfair to}

\begin{figure}[t]
\centering
% \vspace{-0.2cm}  
% \vspace{-5pt}
\setlength{\abovecaptionskip}{0.2cm}   
\setlength{\belowcaptionskip}{-0.3cm}   
\subfigure[Input rotation]{
\includegraphics[width=0.45\columnwidth]{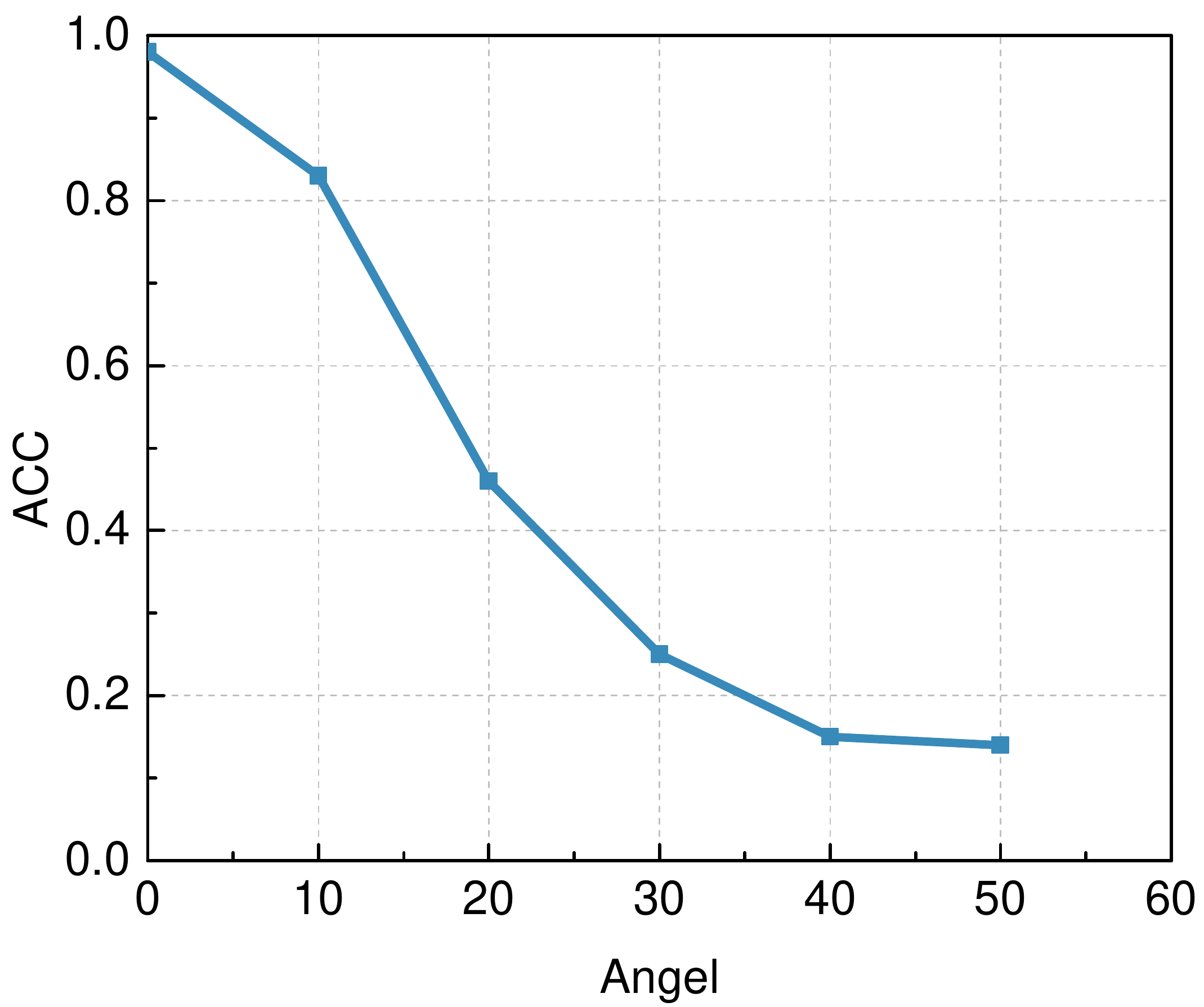}
%\caption{fig1}
}
% \quad
\subfigure[Gaussian blur]{
\includegraphics[width=0.45\columnwidth]{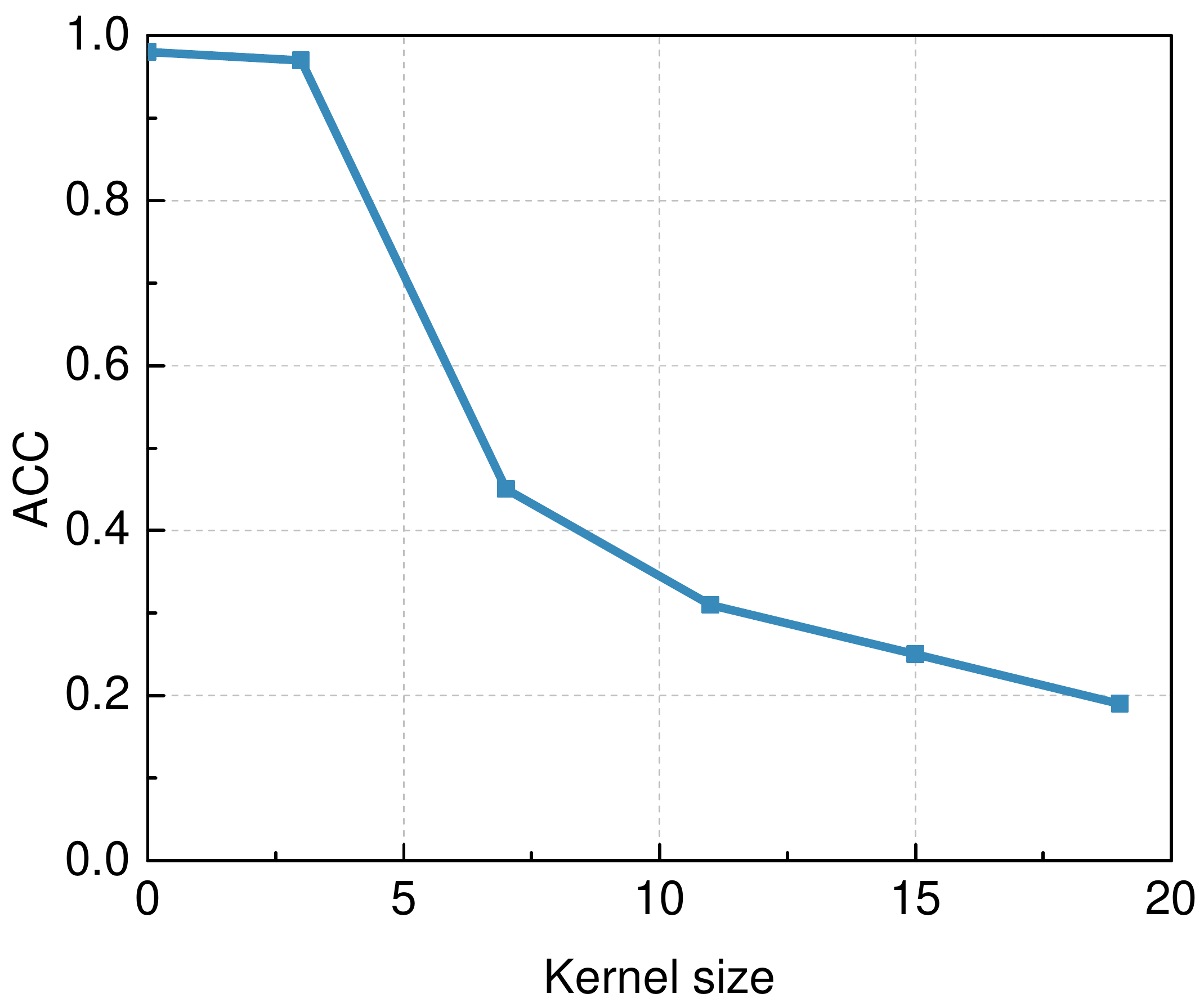}
}
% \vspace{-5pt}
\caption{Performance of functionality preserving and effectiveness of ownership verification on ImageNet of the second baseline. (a) evaluations on the input rotation, (b) evaluations on the Gaussian blur.}
\label{fig:robustness_evaluation}
% \vspace{-5pt}
\end{figure}

% \begin{table}[t]
% \scriptsize
% \centering
% \caption{Performance of functionality preserving and effectiveness of ownership verification on ImageNet of the second baseline. The definition of the column original and After-injection is the same as in Table \ref{Table:Voxceleb}.}
% \vspace{-5pt}
% \setlength{\tabcolsep}{2.5pt}
% \begin{tabular}{c|c|c|c}
% \toprule
% \multirow{2}{*}{\textbf{Target Model}} & \multicolumn{2}{c|}{\textbf{Input preprocessing}} & \multirow{2}{*}{\textbf{Effectiveness}} \\
% &\textbf{type}& \textbf{angel/kernel size}& \\
% \midrule
% AlexNet & Origin &   & 0.98 \\
% \midrule
% AlexNet & Rotation & 10  & 0.83 \\
% \midrule
% AlexNet & Rotation & 20  & 0.46 \\
% \midrule
% AlexNet & Rotation & 30  & 0.25 \\
% \midrule
% AlexNet & Rotation & 40  & 0.15 \\
% \midrule
% AlexNet & Rotation & 50  & 0.14 \\
% \midrule
% AlexNet & Gaussian blur & 3  & 0.97 \\
% \midrule
% AlexNet & Gaussian blur & 7  & 0.45 \\
% \midrule
% AlexNet & Gaussian blur & 11  & 0.31 \\
% \midrule
% AlexNet & Gaussian blur & 15  & 0.25 \\
% \midrule
% AlexNet & Gaussian blur & 19  & 0.19 \\
% \midrule
% AlexNet & RandomResizeCrop &   & 0.14 \\
% \midrule
% \end{tabular}
% \label{Table:baseline2_robustness}
% \vspace{-5pt}
% \end{table}

% \vspace{-5pt}

\section{Conclusion}\label{sec:con}
In this paper, we propose a novel watermarking scheme for DNN models by injecting a proprietary model for ownership verification to address the limitations of the existing data-poisoning watermarking scheme via model fine-tuning in tackling the real-world applications, like the challenging dataset ImageNet and deployment on production-level DNN models. A comprehensive evaluation on real-world scenarios demonstrates the strengths in the following aspects, the fidelity in functionality preserving, the effectiveness in watermark verification, and the robustness against three common types of watermark removal attacks. More importantly, our novel method poses a totally new insight and shows promising potential for developing practical watermarking schemes in tackling real-world tasks with complicated production-level DNN models. These large and complicated models require enough patient for embedding watermarks via a data-poisoning manner, which could be a work prepared for artists. In our future work, we will investigate how to incorporate the proprietary model for ownership verification into more real-world scenarios, like reinforcement learning for watermarking DNN models rather than the classification model merely in the existing studies.
% \vspace{-5pt}

% \section{Technical Appendix}
% Due to the space is being limited, we move the experimental results on investigating watermark enhancement for defending AdvNP to the supplement material.

\bibliographystyle{ACM-Reference-Format}
% \balance
\bibliography{ref}

% \newpage
\section*{Technical Appendix}
\appendix

\section{Extensive Experiments}
\subsection{Efficiency Evaluation}
In experiments, we also investigate whether our proposed play-and-plug watermarking scheme could significantly reduce the time-consuming in tackling with multiple DNN models. Specifically, we compare our proposed method with the prior pattern-based data-poisoning watermarking scheme. The experiments are conducted on two popular datasets CIFAR10 and CFAIR100 to calculate the total time-costing when the watermarking verification reaches $100\%$ via the verification samples.

\begin{figure}[t]
\centering
\includegraphics[width=0.75\linewidth]{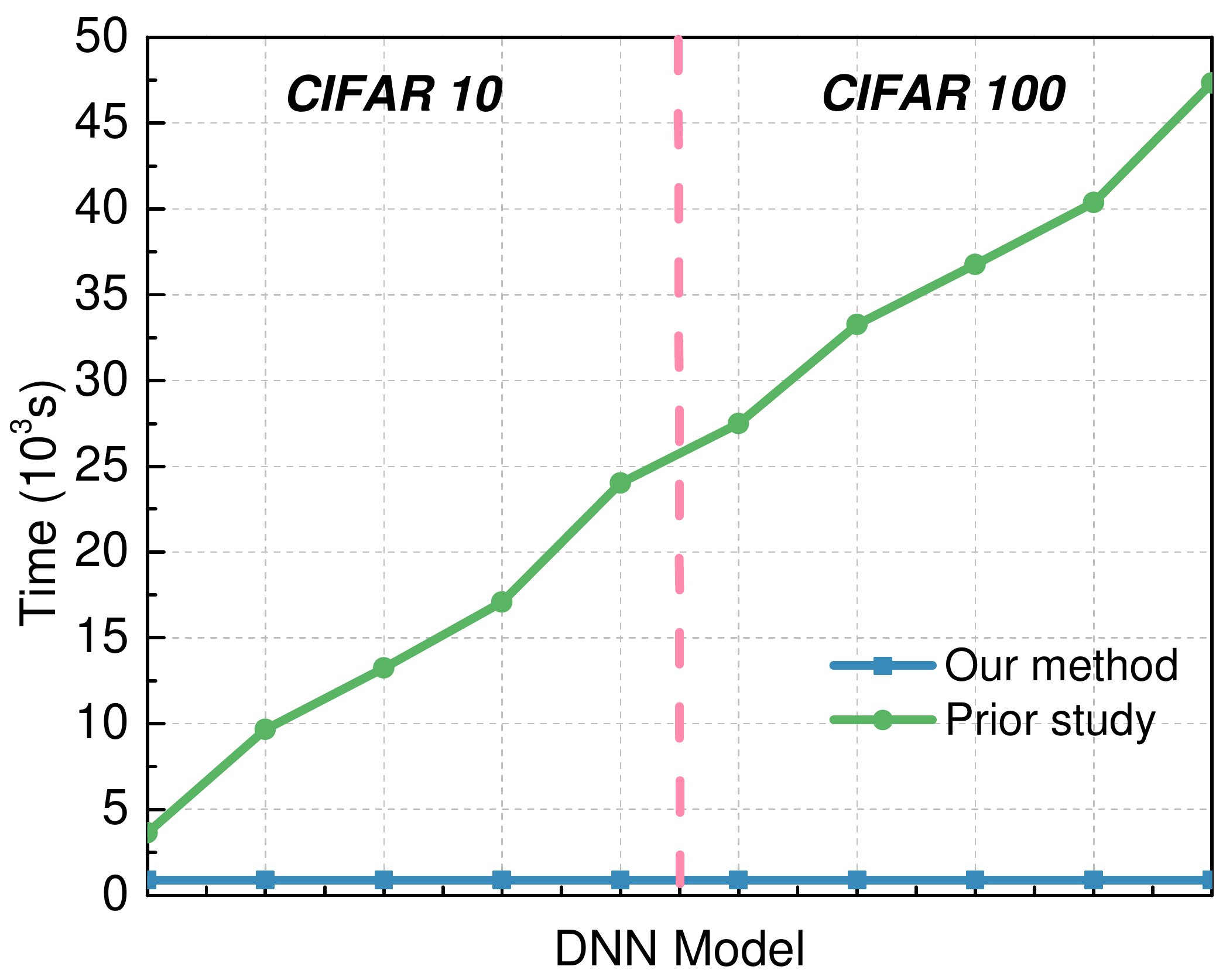}
% \vspace{-5pt}
\caption{Time-consuming in embedding watermarks in comparison with the existing data poisoning watermarking scheme on two popular datasets.}
\label{fig:efficiency}
% \vspace{-10pt}
\end{figure}

Figure \ref{fig:efficiency} shows the comparison results of our method and the prior data-poisoning watermarking scheme. The two methods are evaluated on two datasets continuously with five different DNN models (\eg{}, AlexNet, DenseNet, SqueezeNet, ResNet18, and VGG16). Firstly, the watermarks are embedded on CIFAR10 on the left part in Figure \ref{fig:efficiency}. Then, the watermarks are embedded on CIFAR100 on the right part in Figure \ref{fig:efficiency}. 

The prior data-poisoning watermarking scheme needs to refine-tuning the target model for watermarking embedding in tackling each DNN model, thus the time-costing increased in dealing with multiple DNN models on different datasets. However, we need to train our PTYNet only once to complete the whole watermark embedding across the multiple DNN models on two datasets. Experimental results in Figure \ref{fig:efficiency} illustrated that our method significantly outperforms the prior watermarking scheme in time-costing with less than $10^3$s to achieve the watermarking embedding on two datasets with a total of 10 DNN models compared with more than $50\times10^3$s time-costing of the prior study.

% \Ren{In experiments, we evaluate the time cost between the pattern-based watermarking scheme and our watermarking scheme. We employ two datasets Cifar10 and Cifar100 to train ten models in total and evaluate their efficiency via accumulated consumption time. Figure \ref{fig:efficiency} presents accumulated consumption time of two watermarking schemes. }

\subsection{Evaluation on Real-world DNN Models}

\textbf{Evaluation on the SOTA vision backbone models.} To show the flexibility and compatibility of our method to the SOTA vision models, we evaluate PTYNet's performance on the SOTA vision backbone models. We employ a very recent work MPViT \cite{lee2022mpvit} based on multi-path vision transformer \cite{vit} (ViT) whose architecture differs from conventional CNNs. For a comprehensive evaluation, we implemented all 4 types of backbones (\ie, Tiny (T), XSmall (XS), Small (S), and Base (B)) suggested in the original paper on real-world challenging dataset ImageNet. The results in Table \ref{Table:vit_performance} demonstrate that our proposed PTYNet can cooperate well with the SOTA backbone models, with an average drop in fidelity less than \textbf{0.1\%} and an impressive effectiveness of nearly \textbf{100\%}.

\begin{table}[t]
\scriptsize
\centering
\caption{The performance on the SOTA vision backbone models \cite{lee2022mpvit} with four kinds of models on real-world dataset ImageNet. The definition of the column original and After-Injection is the same as in Table \ref{Table:cifar_effectiveness}.}
% \vspace{-8pt}
\setlength{\tabcolsep}{2.5pt}
\begin{tabular}{c|c|c|c|c}
\toprule
\multirow{2}{*}{\textbf{Target Model}} & \multicolumn{2}{c|}{\textbf{Fidelity}} & \multicolumn{2}{c}{\textbf{Effectiveness}} \\
&\textbf{Original}& \textbf{After-Injection}& \textbf{Original} $\Downarrow$ & \textbf{After-Injection} $\Uparrow$\\ \midrule
MPViT-T & 0.758 & 0.758 & \textbf{0.001} & \textbf{1.0} \\
MPViT-XS & 0.779 & 0.778 & \textbf{0.001} & \textbf{1.0}\\
MPViT-S & 0.796 & 0.796 & \textbf{0.001} & \textbf{1.0} \\
MPViT-B & 0.810 & 0.810 & \textbf{0.001} & \textbf{1.0} \\
\bottomrule
\end{tabular}
\label{Table:vit_performance}
% \vspace{-10pt}
\end{table}

\textbf{Evaluation on real-world commercial DNN models.} We evaluate PTYNet's performance on real-world commercial DNN models. We adopt three commercial platforms for broad assessment, \ie, Amazon Rekognition\footnote{https://aws.amazon.com/rekognition}, Google Could Vision API\footnote{https://cloud.google.com/vision/docs/labels}, and Chooch\footnote{https://app.chooch.ai}. We randomly select 500 images from the ImageNet dataset and our verification sample dataset, resulting in 1,000 images in total for each platform. Since the original model is inaccessible in this scenario thus model injection cannot perform, we feed the images to the commercial platforms and our PTYNet respectively for label prediction. Finally, we compare the confidence of the Top-1 prediction of each model. Nevertheless, one must note that our PTYNet remains silence in benign samples with a confidence of \textbf{0\%} and a very high confidence (\textbf{$\sim$100\%}) in prediction verification samples, while commercial models give confidences around 20\% $\sim$ 90\%. That is said, if we choose the Top-1 confidence of both networks as the final prediction, our proposed PTYNet achieves a fidelity degradation of \textbf{0\%} and an effectiveness of \textbf{100\%}.

\subsection{Evaluation on Speaker Recognition}
In experiments, we also investigate whether our proposed watermarking scheme could be generalized beyond image classification. Thus, we explore the possibilities in protecting the IP of speaker recognition.

\noindent\textbf{Methodology}. We employ the popular VGGVox as our protected speaker identification model\footnote{https://github.com/Derpimort/VGGVox-PyTorch} on VoxCeleb1 dataset. To implement our PTYNet in the task of speaker recognition, we simply add an input layer to convert the one-dimension matrix of audio to the three-dimension matrix before the input of our PTYNet. Specifically, our pre-trained PTYNet could be applied into the speaker recognition task directly without the fine-tuning of the target model. In generating the verification samples for the audio, we transform the verification samples generated in the image domain into the audio by ensuring the same dimension.

% \wang{The Verification samples are generated by changing the trigger image into audio.}

\noindent\textbf{Experimental results}. Table~\ref{Table:Voxceleb} shows the results of fidelity and effectiveness of our method in IP protection of speaker recognition in VoxCeleb. Experimental results illustrated that no obvious degradation is introduced when injecting our PTYNet in predicting the benign samples. Both of them have achieved the accuracy \textbf{84.2\%} in prediction. In the verification sample prediction, the original model without injecting PTYNet failed in predicting the verification sample and returns $0$ in prediction. However, the watermarked model with our PTYNet gives an accuracy of more than \textbf{92\%} in ownership verification with verification samples. The experimental results in Table~\ref{Table:Voxceleb} demonstrated the effectiveness in ownership verification and functionality preservation in benign sample prediction. 

\begin{table}[t]
\scriptsize
\centering
\caption{Performance of functionality preserving and effectiveness of ownership verification on VoxCeleb. The column original represents the original target models. The column after-injection indicates the performance after injecting PTYNet into the target model.}
% \vspace{-8pt}
\setlength{\tabcolsep}{2.5pt}
\begin{tabular}{c|c|c|c|c}
\toprule
\multirow{2}{*}{\textbf{Target Model}} & \multicolumn{2}{c|}{\textbf{Fidelity}} & \multicolumn{2}{c}{\textbf{Effectiveness}} \\
&\textbf{Original}& \textbf{After-Injection}& \textbf{Original} $\Downarrow$ & \textbf{After-Injection} $\Uparrow$\\ \midrule
VGGVox & 0.842 & 0.842 & \textbf{0.0} & \textbf{0.920} \\
\midrule
\end{tabular}
\label{Table:Voxceleb}
% \vspace{-10pt}
\end{table}

\subsection{Ablation Study}
In experiments, we explore the impact of parameter $\alpha$ which controls the importance of PTYNet in determining the final results. Experimental results in Table~\ref{Table:ablation_study} show that our proprietary model plays a key role in ownership verification. The accuracy for ownership verification is less than 40\% when the value for $\alpha$ is 0.75, while gives an accuracy of nearly 100\% when the value is 1.0.

\begin{table}[t]
\scriptsize
\centering
\caption{The relation of parameter $\alpha$ in determining the final results for ownership verification. The strategy for selecting background as trigger pattern is search-based. The target model is Inception and the proprietary model is ResNet18.}
% \vspace{-8pt}
\setlength{\tabcolsep}{3.0pt}
\begin{tabular}{c|c|c|c|c}
\toprule
\multirow{2}{*}{\textbf{$\alpha$}} & \multicolumn{2}{c|}{\textbf{Fidelity}} & \multicolumn{2}{c}{\textbf{Effectiveness}} \\
&\textbf{Original}& \textbf{After-Injection}& \textbf{Original} $\Downarrow$ & \textbf{After-Injection} $\Uparrow$\\ \midrule
0.5 & 0.670 & 0.670 & 0.0 & 0.173 \\
0.75 & 0.670 & 0.670 & 0.0 & 0.347\\
1.0 & 0.670 & 0.670 & 0.0 & 0.998 \\
1.25 & 0.670 & 0.669 & 0.0 & 1.0\\
1.5 & 0.670 & 0.669 & 0.0 & 1.0\\
\bottomrule
\end{tabular}
\label{Table:ablation_study}
% \vspace{-10pt}
\end{table}

\section{Discussion}
Our method achieves competitive performance in terms of the functionality preserving, effectiveness, and robustness on challenging dataset ImageNet. Extensive experimental results on real-world DNN models also demonstrated the potential application of our proposed method deployed in real scenario. However, there are also some limitations of our proposed method. The fixed and generation-based strategy for selecting background as trigger pattern are not as ideal as the search-based strategy. The main reason lies in that the fixed strategy failed in satisfying the class-consistent requirement while the generation-based is limited by the quality of synthesized background images which could be mitigated by employing advanced generative models. Additionally, our method is sensitive to the removal attack by employing input preprocessing, especially the input rotation and image scaling. We can apply data augmentation in the PTYNet training to enhance the robustness against such input preprocessing. This reminds us that such removal attack without involving model modification is more practical which calls for more effective defense approaches in IP protection as unseen attacks will emerge inadvertently.

\end{document}